\documentstyle[12pt,procsla,cite,psfig,axodraw]{article}

\makeatletter

\newtoks\@stequation

\def\subequations{\refstepcounter{equation}%
  \edef\@savedequation{\the\c@equation}%
  \@stequation=\expandafter{\theequation}
  \edef\@savedtheequation{\the\@stequation}
  \edef\oldtheequation{\theequation}%
  \setcounter{equation}{0}%
  \def\theequation{\oldtheequation\alph{equation}}}

\def\endsubequations{%
  \ifnum\c@equation < 2 \@warning{Only \the\c@equation\space subequation
    used in equation \@savedequation}\fi
  \setcounter{equation}{\@savedequation}%
  \@stequation=\expandafter{\@savedtheequation}%
  \edef\theequation{\the\@stequation}%
  \global\@ignoretrue}

\def\eqnarray{\stepcounter{equation}\let\@currentlabel\theequation
\global\@eqnswtrue\m@th
\global\@eqcnt\z@\tabskip\@centering\let\\\@eqncr
$$\halign to\displaywidth\bgroup\@eqnsel\hskip\@centering
     $\displaystyle\tabskip\z@{##}$&\global\@eqcnt\@ne
      \hfil$\;{##}\;$\hfil
     &\global\@eqcnt\tw@ $\displaystyle\tabskip\z@{##}$\hfil
   \tabskip\@centering&\llap{##}\tabskip\z@\cr}

\def\dlinepattern#1#2{%
\ifdim#2<#1
   \errmessage{the 1st argument is less than the 2nd argument.}%
\else
   \gdef\dline@solid{#1}\gdef\dline@period{#2}%
\fi}

\def\dline#1{\@dline[#1]}
\def\@dline[#1-#2]{\noalign{\global\@dla#1\relax
\global\advance\@dla\m@ne
\ifnum\@dla>\z@\global\let\@gtempa\@dlinea\else
  \global\let\@gtempa\@dlineb\fi
\global\@dlb#2\relax
\global\advance\@dlb-\@dla}\@gtempa
\noalign{\vskip-\arrayrulewidth}}

\def\@dlinea{\multispan\@dla&\multispan\@dlb
\unskip\cleaders\hbox to \dline@period
{\hss\rule{\dline@solid}{\arrayrulewidth}\hss}\hfill\cr}

\newcount\@dla
\newcount\@dlb

\def\@dlineb{\multispan\@dlb
\unskip\cleaders\hbox to \dline@period
{\hss\rule{\dline@solid}{\arrayrulewidth}\hss}\hfill\cr}

\dlinepattern{2pt}{5pt}

\makeatother

\newcommand\textfrac[2]{{\textstyle\frac{#1}{#2}}}

\def\simgt{\rlap{\lower 3.5 pt \hbox{$\mathchar \sim$}}%
           \raise 1pt \hbox {$>$}}
\def\simlt{\rlap{\lower 3.5 pt \hbox{$\mathchar \sim$}}%
           \raise 1pt \hbox {$<$}}

\def\gev{{\,\rm GeV}}

\def\tev{{\,\rm TeV}}

\def\msbar{\overline{\rm MS }}	
\def\to{\rightarrow}

\def\mz{m_Z^{}}
\def\mw{m_W^{}}
\def\mh{m_H^{}}
\def\mmv{m_V^2}
\def\mmz{m_Z^2}
\def\mmw{m_W^2}
\def\mmh{m_H^2}
\def\ebar{\bar{e}}
\def\sbar{\bar{s}}
\def\cbar{\bar{c}}
\def\gzbar{\bar{g}_Z}
\def\gwbar{\bar{g}_W}
\def\ehat{\hat{e}}
\def\shat{\hat{s}}
\def\chat{\hat{c}}
\def\gzhat{\hat{g}_Z}
\def\ghat{\hat{g}}

\def\pibar{\overline{\Pi}}

\def\delb{\bar{\delta}_{b}}
\def\delg{\bar{\delta}_{G}^{}}
\def\zbb{Zb_L^{}b_L^{}}

\def \ibid {{\it ibid}.}
\def \etal {{\it et al}.}

\newcommand {\MPL}[3]	{{\it Mod.\ Phys.\ Lett.} {\bf #1} (#2) #3}

\newcommand {\NPB}[3]	{{\it Nucl.\ Phys.} {\bf B#1} (#2) #3}
\newcommand {\PL}[3]	{{\it Phys.\ Lett.} {\bf #1} (#2) #3}
\newcommand {\PLB}[3]	{{\it Phys.\ Lett.} {\bf B#1} (#2) #3}
\newcommand {\PR}[3]	{{\it Phys.\ Rev.} {\bf #1} (#2) #3}
\newcommand {\PRD}[3]	{{\it Phys.\ Rev.} {\bf D#1} (#2) #3}
\newcommand {\PRL}[3]	{{\it Phys.\ Rev.\ Lett.} {\bf#1} (#2) #3}

\newcommand {\ZP}[3]	{{\it Z.~Phys.} {\bf #1} (#2) #3}
\newcommand {\ZPC}[3]	{{\it Z.~Phys.} {\bf C#1} (#2) #3}

\newcommand {\both}[1]	{\relax\ifmmode#1\else$#1$\fi\relax}

\newcommand {\bea}	{\begin{eqnarray}}
\newcommand {\eea}	{\end{eqnarray}}
\newcommand {\be}	{\begin{equation}}
\newcommand {\ee}	{\end{equation}}
\newcommand {\bsub}	{\begin{subequations}}
\newcommand {\esub}	{\end{subequations}}

\begin{document}
\thispagestyle{empty}

\vspace*{-2cm}
\begin{flushright}
\begin{tabular}{l}
KEK--TH--461 \\[-1mm]
KEK preprint 95--184 \\[-1mm]
December 1995
\end{tabular}
\end{flushright}
\vspace*{20mm}

\centerline{\normalsize\bf IMPLICATIONS OF PRECISION ELECTROWEAK DATA }
\baselineskip=22pt

\centerline{\footnotesize KAORU HAGIWARA }
\baselineskip=13pt
\centerline{\footnotesize\it Theory Group, KEK }
\baselineskip=12pt
\centerline{\footnotesize\it Oho, Tsukuba, Ibaraki 305, Japan }

%

\vspace*{0.9cm}
\abstracts{
There are two aspects to the 1995 summer update of the combined preliminary
electroweak data from LEP and SLC.  On the one hand, agreement between
experiments and the Standard Model (SM) has improved for the line-shape
and the asymmetry data.  The $\tau$ widths and asymmetries are now consistent
with $e$--$\mu$--$\tau$ universality, and all the asymmetry data including
the left-right asymmetry from SLC are consistent with the SM (16\%CL).
On the other hand, a discrepancy between experiments and SM predictions
is sharpened for two observables, $R_b$ and $R_c$, where $R_q$ is the
partial $Z$ boson width ratio $\Gamma_q/\Gamma_h$.
$R_b$ is 3\% larger (3.7$\sigma$) and $R_c$ is 11\% smaller (2.5$\sigma$)
than the SM predictions.  When combined, the SM is ruled out at the 99.99\%CL
for $m_t>170$GeV.
It is difficult to interpret the $11\pm 4$\% deficit of $R_c$, since
if we allow only $\Gamma_b$ and $\Gamma_c$ to deviate from the SM then
the precisely measured ratio $R_h=\Gamma_h/\Gamma_\ell$ forces the QCD
coupling to be
$\alpha_s\equiv \alpha_s(m_Z)_{\overline{\rm MS}}=0.185\pm 0.041$,
which is uncomfortably large.
The data can be consistent with the prefered $\alpha_s$
($0.10<\alpha_s<0.13$) only if the sum $\Gamma_h=\Sigma_q \Gamma_q$
does not deviate significantly from the SM prediction.
Possible experimental causes for the under-estimation of $R_c$ are discussed.
By assuming the SM value for $R_c$, the discrepancy in $R_b$ decreases
to 2\% (3$\sigma$).
The double tagging technique used for the $R_b$ measurements is critically
reviewed.  A few theoretical models that can explain large $R_b$ and
small $\alpha_s$ ($= 0.104\pm 0.008$) are discussed.
If the QCD coupling $\alpha_s$ is allowed to be fitted by the data,
the standard $S$, $T$, $U$ analysis for a new physics search in the
gauge-boson propagator corrections does not suffer from the $R_b$
and $R_c$ crisis.  No signal of new physics is found in the $S$, $T$, $U$
analysis once the SM contributions with $m_t \sim 175$GeV has
been accounted for.
The naive QCD-like technicolor model is now ruled out at the 99\%CL
even for the minimal model with ${\rm SU(2)_{TC}}$.
By assuming that no new physics effect is significant in the electroweak
observables, we obtain constraints on $m_t$ and $m_H$ as a function
of $\alpha_s$ and $\bar{\alpha}(m_Z^2)$, the QED coupling constant at
the $m_Z$ scale.
A lighter Higgs boson $m_H\simlt 200$GeV is prefered if $m_t<170$GeV.
The controversy in $\bar{\alpha}(m_Z^2)$ is overcome.
However, further improvements in our knowledge of its numerical value
is essential in order for the electroweak precision experiments to be
sensitive to new physics effects in quantum corrections.
}
\normalsize\baselineskip=15pt
\setcounter{footnote}{0}
\renewcommand{\thefootnote}{\alph{footnote}}

\vspace*{20mm}
\begin{center}
{\it Talk presented at XVII International Symposium on Lepton and Photon
Interactions at High Energies, 10--15 August 1995, Beijing, China }
\end{center}

\vfil
\newpage
\setcounter{page}{1}
\def\gzbarrunning{%
  \begin{equation}
     \frac{1}{\gzbar^2(\mmz)}\approx\frac{1}{\gzbar^2(0)}
       -0.02390  +\frac{2.41}{m_t^2} +\frac{1.73}{\mmh}
   \label{gzbarrunning}
  \end{equation}
}

\def\fitofgzbsb{%
  \begin{subequations}
   \label{fitofgzbsb}
  \begin{eqnarray}
     & &
     \left.
     \begin{array}{ll}
     \gzbar^2(\mmz) &\!\!= 0.55556
      -0.00049\,\textfrac{\alpha_s' -0.1081}{0.0043}
       \pm 0.00072
      \\[1mm]
     \sbar^2(\mmz)  &\!\!= 0.23041
      +0.00004\,\textfrac{\alpha_s' -0.1081}{0.0043}
       \pm 0.00029
     \end{array}
    \right\}\,\,
   \rho_{\rm corr} = 0.23,
   \\
   & & \quad
   \chi^2_{\rm min} =  15.6
         +\biggl(\frac{\alpha_s' -0.1081}{0.0043}\biggr)^2
         +\biggl(\frac{\delb-0.0025}{0.0042}\biggr)^2\,,
   \label{fitofgzbsbchisq}
  \end{eqnarray}
  \end{subequations}
}

\def\fitofmw{%
  \begin{equation}
     \label{fitofmw}
	\gwbar^2(0) = 0.4227\pm 0.0017\,,
  \end{equation}
}%



\def\fitofstu{%
  \begin{subequations}
     \label{fitofstu}
  \begin{eqnarray}
  &&\!\! \left.
    \begin{array}{c@{}r@{}r@{}r@{}r}
       S =&   -0.42 &
              -0.059\,\frac{\alpha_s'-0.1093}{0.0042} &
              +0.06\,\frac{\delta_\alpha-0.03}{0.09} &
              \pm 0.15
         \\[1mm]
       T =&    0.57 &
              -0.104\,\frac{\alpha_s'-0.1093}{0.0042} & &
              \pm 0.17
         \\[1mm]
       U =&    0.16 &
              +0.079\,\frac{\alpha_s'-0.1093}{0.0042} &
              +0.02\,\frac{\delta_\alpha-0.03}{0.09} &
              \pm 0.49 \\[1mm]
    \end{array}
    \right\} \;
   \rho_{\rm corr} = \left(
      \begin{array}{rrr}
           1    &  0.86 & -0.10  \\[1mm]
                &  1    & -0.20  \\[1mm]
                &       &  1     \\[1mm]
      \end{array}
      \right), \,\,\,
   \qquad \label{fitofstu_mean}
   \\
   & & \chi^2_{\rm min} =  20.4
         +\biggl(\frac{\alpha_s' -0.1093}{0.0042}\biggr)^2
         +\biggl(\frac{\delb-0.0025}{0.0042}\biggr)^2\,.
     \label{chisqofstu}
  \end{eqnarray}
  \end{subequations}
}%

\def\dataofnqccfrorig{%
\begin{equation}
   K = 0.5626 \pm 0.0025\,\mbox{(stat)} \pm 0.0036\,\mbox{(sys)}
              \pm 0.0028\,\mbox{(model)}\pm 0.0029\,\mbox{($m_c$)}.
   \label{dataofnqccfrorig}
\end{equation}
}

\def\dataofnqccfr{%
\begin{equation}
   K = 0.5626 \pm 0.0060\,\mbox{(stat,sys,model,$m_c$)}
   \label{dataofnqccfr}
\end{equation}
}

\def\fitofnqccfr{%
\begin{eqnarray}
  \sbar^2(0)
             = 0.2421 + 1.987[\gzbar^2(0)-0.5486] \pm 0.0058\,.
   \label{fitofnqccfr}
\end{eqnarray}
}

\def\fitofnqfh{%
  \begin{subequations}
   \label{fitofnqfh}
  \begin{eqnarray}
     & &
     \left.
     \begin{array}{ll}
     \gzbar^2(0) &= 0.5454^{+0.0076}_{-0.0082}
      \\[1mm]
     \sbar^2(0)  &= 0.2419^{+0.0130}_{-0.0142}
     \end{array}
    \right\}\quad
   \rho_{\rm corr} = 0.916,
   \\
   & & \quad
   \chi^2_{\rm min} =  0.13\,.
   \label{fitofnqfhchisq}
  \end{eqnarray}
  \end{subequations}
}

\def\fitofnq{%
  \begin{subequations}
   \label{fitofnq}
  \begin{eqnarray}
     & &
     \left.
     \begin{array}{ll}
     \gzbar^2(0) &= 0.5476^{+0.0070}_{-0.0076}
      \\[1mm]
     \sbar^2(0)  &= 0.2429^{+0.0128}_{-0.0140}
     \end{array}
    \right\}\quad
   \rho_{\rm corr} = 0.955,
   \\
   & & \quad
   \chi^2_{\rm min} =  0.7
   \label{fitofnqchisq}
  \end{eqnarray}
  \end{subequations}
}

\def\fitoflenc{%
  \begin{subequations}
   \label{fitoflenc}
  \begin{eqnarray}
     & &
     \left.
     \begin{array}{ll}
     \gzbar^2(0) &= 0.5441 \pm 0.0029
      \\[1mm]
     \sbar^2(0)  &= 0.2362 \pm 0.0044
     \end{array}
    \right\}\quad
   \rho_{\rm corr} = 0.70,
   \\
   & & \quad
   \chi^2_{\rm min} =  2.7\,.
   \label{fitoflencchisq}
  \end{eqnarray}
  \end{subequations}
}

\def\fitoflencatmz{%
  \begin{subequations}
   \label{fitoflencatmz}
  \begin{eqnarray}
     & &
     \left.
     \begin{array}{ll}
     \gzbar^2(\mmz) &= 0.5512 \pm 0.0030
      \\[1mm]
     \sbar^2(\mmz)  &= 0.2277 \pm 0.0047
     \end{array}
    \right\}\quad
   \rho_{\rm corr} = 0.70,
   \\
   & & \quad
   \chi^2_{\rm min} =  2.7\,.
   \label{fitoflencatmzchisq}
  \end{eqnarray}
  \end{subequations}
}

\def\fitoflenchhkm{%
  \begin{subequations}
   \label{fitoflenchhkm}
  \begin{eqnarray}
     & &
     \left.
     \begin{array}{ll}
     \gzbar^2(0) &= 0.5425 \pm 0.0035
      \\[1mm]
     \sbar^2(0)  &= 0.2365 \pm 0.0045
     \end{array}
    \right\}\quad
   \rho_{\rm corr} = 0.53,
   \\
   & & \quad
   \chi^2_{\rm min} =  2.0
   \label{fitoflenchhkmchisq}
  \end{eqnarray}
  \end{subequations}
}
\def\chisqsm{
  \begin{subequations}
  \label{total_chisqsm}
  \begin{eqnarray}
     \chi^2_{\rm SM}(m_t,\mh,\alpha_s,\delta_\alpha)
     &=& \biggl(\frac{m_t -\langle m_t\rangle}{\Delta m_t} \biggr)^2
         +\chi^2_{H}(\mh,\alpha_s,\delta_\alpha)\,,
     \label{chisqsm}
  \end{eqnarray}
}
%
%
\def\fitofmt{%
  \begin{eqnarray}
     \label{fitofmt}
    \langle m_t \rangle &=& 163.3 +13.1\, \ln\frac{\mh}{100}
                            +0.8 \, \ln^2\frac{\mh}{100}
           -3.1\,\biggl(\frac{\alpha_s-0.12}{0.01}\biggr)
           -4.9\,\biggl(\frac{\delta_\alpha-0.03}{0.09}\biggr),\qquad
     \label{fitofmtbest}
     \\
     \Delta m_t   &=& 6.7 - 0.07\, \ln \frac{\mh}{100}
          - \Bigl(0.018 -0.003\,\ln \frac{\mh}{100}\Bigr)\,
            \frac{m_t -175}{10}\,,
     \label{fitofmterror}
  \end{eqnarray}
}
%
%
\def\chisqhsm{%
  \begin{eqnarray}
     \chi^2_H(\mh,\alpha_s,\delta_\alpha) &=&
      27.0 + \biggl(\frac{\delta_\alpha -0.44}{0.23} \biggr)^2
          + \biggl(\frac{\alpha_s-0.1222+0.0014\,\delta_\alpha}{0.0036}
            \biggr)^2
     \nonumber \\
     && \!\!\!\!\!
          - \biggl(\frac{\alpha_s-0.1470+0.052\,\delta_\alpha}{0.0087}
                   \biggr) \ln \frac{\mh}{100}
          - \biggl(\frac{\alpha_s-0.1315}{0.0174}\biggr)
                   \ln^2 \frac{\mh}{100} \,.
     \nonumber \\ &&   \label{chisqhsm}
  \end{eqnarray}
  \end{subequations}
}

\centerline{\normalsize\bf IMPLICATIONS OF PRECISION ELECTROWEAK DATA }
\baselineskip=22pt

\centerline{\footnotesize KAORU HAGIWARA }
\baselineskip=13pt
\centerline{\footnotesize\it Theory Group, KEK }
\baselineskip=12pt
\centerline{\footnotesize\it Oho, Tsukuba, Ibaraki 305, Japan }

\vspace*{0.9cm}
\abstracts{
There are two aspects to the 1995 summer update of the combined preliminary
electroweak data from LEP and SLC.  On the one hand, agreement between
experiments and the Standard Model (SM) has improved for the line-shape
and the asymmetry data.  The $\tau$ widths and asymmetries are now consistent
with $e$--$\mu$--$\tau$ universality, and all the asymmetry data including
the left-right asymmetry from SLC are consistent with the SM (16\%CL).
On the other hand, a discrepancy between experiments and SM predictions
is sharpened for two observables, $R_b$ and $R_c$, where $R_q$ is the
partial $Z$ boson width ratio $\Gamma_q/\Gamma_h$.
$R_b$ is 3\% larger (3.7$\sigma$) and $R_c$ is 11\% smaller (2.5$\sigma$)
than the SM predictions.  When combined, the SM is ruled out at the 99.99\%CL
for $m_t>170$GeV.
It is difficult to interpret the $11\pm 4$\% deficit of $R_c$, since
if we allow only $\Gamma_b$ and $\Gamma_c$ to deviate from the SM then
the precisely measured ratio $R_h=\Gamma_h/\Gamma_\ell$ forces the QCD
coupling to be
$\alpha_s\equiv \alpha_s(m_Z)_{\overline{\rm MS}}=0.185\pm 0.041$,
which is uncomfortably large.
The data can be consistent with the prefered $\alpha_s$
($0.10<\alpha_s<0.13$) only if the sum $\Gamma_h=\Sigma_q \Gamma_q$
does not deviate significantly from the SM prediction.
Possible experimental causes for the under-estimation of $R_c$ are discussed.
By assuming the SM value for $R_c$, the discrepancy in $R_b$ decreases
to 2\% (3$\sigma$).
The double tagging technique used for the $R_b$ measurements is critically
reviewed.  A few theoretical models that can explain large $R_b$ and
small $\alpha_s$ ($= 0.104\pm 0.008$) are discussed.
If the QCD coupling $\alpha_s$ is allowed to be fitted by the data,
the standard $S$, $T$, $U$ analysis for a new physics search in the
gauge-boson propagator corrections does not suffer from the $R_b$
and $R_c$ crisis.  No signal of new physics is found in the $S$, $T$, $U$
analysis once the SM contributions with $m_t \sim 175$GeV has
been accounted for.
The naive QCD-like technicolor model is now ruled out at the 99\%CL
even for the minimal model with ${\rm SU(2)_{TC}}$.
By assuming that no new physics effect is significant in the electroweak
observables, we obtain constraints on $m_t$ and $m_H$ as a function
of $\alpha_s$ and $\bar{\alpha}(m_Z^2)$, the QED coupling constant at
the $m_Z$ scale.
A lighter Higgs boson $m_H\simlt 200$GeV is prefered if $m_t<170$GeV.
The controversy in $\bar{\alpha}(m_Z^2)$ is overcome.
However, further improvements in our knowledge of its numerical value
is essential in order for the electroweak precision experiments to be
sensitive to new physics effects in quantum corrections.
}
\normalsize\baselineskip=15pt
\setcounter{footnote}{0}
\renewcommand{\thefootnote}{\alph{footnote}}

\section{Introduction}

Despite the overwhelming success of the Standard Model (SM) of
the electroweak interactions when confronted with experimental
observations, there has been a strong and steady belief that
the SM is merely an effective low energy description of a more
fundamental theory.
Moreover, naturalness of the dynamics of the electroweak gauge
symmetry breakdown suggests that the energy scale of new physics
beyond the SM should lie below or at $\sim 1$~TeV.
This is so whether its last missing ingredient, the Higgs boson,
exists or not.
It has therefore been hoped that hints of new physics beyond
the SM may be found as quantum effects affecting precision
electroweak observables.

In response to such general expectations, the experimental
accuracy of the electroweak measurements has steadily been
improved in the past several years,
reaching the $10^{-5}$ level for $\mz$,
a few $\times 10^{-3}$ level for the total and
some of the partial $Z$ widths,
and $10^{-2}$ level for the asymmetries at LEP and SLC.
Because of partial cancellation in the observable asymmetries
at LEP and SLC, their measurements at the $10^{-2}$ level
determine the effective electroweak mixing parameter
$\sin^2\theta_W$ at the $10^{-3}$ level.
Therefore, by choosing the fine structure constant, $\alpha$,
the muon-decay constant $G_F$, and $\mz$ as the three inputs
whose measurement error is negligibly small,
we can test the predictions of the SM at
a few $\times 10^{-3}$ level.
Accuracy of experiments has now reached the level where
new physics contributions to quantum corrections
can be probed.

The precision electroweak measurements which were
reported as preliminary results for the 1995 Summer
Conferences\cite{lepewwg9502,lephf9502} are, however,
characterized by the following two conflicting aspects.

On the one hand, {\em all} the observables that were
measured at a few $\times 10^{-3}$ level are in perfect
agreement with the predictions of the SM.
We find no hint of new physics there,
and the data are starting to constrain the Higgs boson mass,
$\mh$, in the minimal SM framework provided the top-quark mass,
$m_t$, will be known accurately in the future.
The precision of these tests has already reached the
level where the present uncertainty of
$0.7\times 10^{-3}$\cite{eidjeg95}
in the running QED coupling constant at the $\mz$ scale,
$\bar{\alpha}(\mmz)$, is no longer negligible as compared
to the other experimental errors which have been attained
at LEP and SLC.

On the other hand, significant deviations from the SM predictions
are found for the two ratios of the $Z$ partial widths,
$R_b=\Gamma(Z\to \mbox{`$b\bar{b}$'} )/\Gamma(Z\to \rm{hadrons})$ and
$R_c=\Gamma(Z\to \mbox{`$c\bar{c}$'} )/\Gamma(Z\to \rm{hadrons})$,
which are measured at the 1\% and 4\% level, respectively.
The disagreements are significant, more than 3-$\sigma$
for $R_b$ and more than 2-$\sigma$ for $R_c$.
When combined, the SM can be ruled out at 99.99\%CL
for $m_t > 170$~GeV\cite{renton}.

Our task is hence to try to find a consistent picture of
electroweak physics that can accommodate simultaneously
the above two features of the most recent precision experiments.
I would like to report difficulties that I encountered
during this course of studies.

The report is organized as follows.
In section 2
we summarize the preliminary results\cite{lepewwg9502,lephf9502}
of the electroweak measurements at LEP and SLC, reported
at this Symposium\cite{renton}.
These data are then compared with the SM predictions\cite{hhkm},
and a few remarkable features are pointed out.
In section 3
we discuss the nature of the $R_b$ and $R_c$ crisis in detail,
and we show that its resolution is intimately related to
the possible problem of the magnitude of the strong coupling
constant, $\alpha_s \equiv \alpha_s(\mz )_{\msbar}$.
In particular, it is pointed out that the data on $R_c$ imply
too large an $\alpha_s$ in conflict with
its recent measurements\cite{bethke95}, if the $Z$ partial
widths into light quarks ($u$, $d$, $s$) were consistent with
the SM predictions.
If, on the other hand, we assume the SM value for $R_c$,
then the data on $R_b$ implies $\alpha_s \sim 0.11$
which is consistent with the estimates from the low-energy
experiments.
A few theoretical ideas that could explain the $R_b$ data
are briefly discussed.
In section 4
we perform the comprehensive fit to all the electroweak data
by allowing the three parameters\cite{stu} $S$, $T$, $U$
characterizing possible new physics contributions through
the electroweak gauge-boson propagator corrections to vary.
Although we assume the SM value for $R_c$ in this analysis,
the effects of the new $R_b$ data on this general fit are
studied carefully.
The simple QCD-like Techni-Color (TC) model is ruled out at
the 99\%CL even for the minimal model despite the $R_b$ data
if we allow $\alpha_s$ to be varied in the fit.
The uncertainty in the running QED coupling constant at
the $\mz$ scale, $\bar{\alpha}(\mmz)$, is shown as the
serious limiting factor for future improvements in the
measurement of the $S$ parameter.
In section 5
we perform the minimal SM fit to all the electroweak data.
Here, despite the $R_b$ and $R_c$ problem, all the electroweak
data taken together is consistent with the SM at
a few to several \%CL for prefered ranges of ($m_t$, $\mh$)
and $\alpha_s$.
This is a consequence of the excellent agreement between the
SM predictions and the rest of the precision data.
We show constraints on ($m_t$, $\mh$) as functions of
$\alpha_s$ and $\bar{\alpha}(\mmz)$.
Improved numerical precision for the above two coupling
constants is essential to improve the constraint on
$\mh$ in the minimal SM, and hence to detect new physics
effects in quantum corrections.
Finally, the quantitative significance of the fermionic and
the bosonic radiative corrections is discussed briefly.
Section 6 summarizes our findings.

\section{%
Precision Electroweak Data }

Table~1 summarizes the results of
the LEP Electroweak Working Group\cite{lepewwg9502,lephf9502},
which are obtained by combining preliminary electroweak data from
LEP, SLC and Tevatron.
Correlation matrices among the errors of the line-shape parameters
and the heavy-quark parameters are given in Tables~2 and 3,
respectively.
The errors and their correlations were obtained by combining
statistical and systematic errors of individual experiments.
All the numerical results presented in this report are obtained
by using the data in Tables~1--3, unless otherwise stated.

Also shown in Table~1 are the SM predictions\cite{hhkm} for
$m_t=175$~GeV, $\mh=100$~GeV, $\alpha_s(\mz )=0.12$ and
$1/\bar{\alpha}(\mmz )=128.75$.
We will discuss implications of the QCD and QED running coupling
strengths in sections 3 and 4, respectively.
The right-most column gives the difference between the mean of
the data and the corresponding SM prediction in units of the
experimental error.
The data and the SM predictions agree well for most of the
observables except for the two ratios $R_b$ and $R_c$
which are, respectively, the partial $Z$ decay widths into
$b\bar{b}$- and $c\bar{c}$-initiated hadronic states,
$\Gamma_b$ and $\Gamma_c$, divided by the $Z$ hadronic decay
width $\Gamma_h$.
$R_b$ is larger than the SM prediction by 3.7-$\sigma$,
whereas $R_c$ is smaller than the prediction by 2.5-$\sigma$.
The trends of larger $R_b$ and smaller $R_c$ existed in the
combined data for the past few years, but their significance
grew considerably in the updated data.

\begin{table}[t]
\begin{center}
\tcaption{%
Preliminary electroweak results from LEP, SLC and Tevatron
for the 1995 summer conferences\protect\cite{lepewwg9502,lephf9502}.
The SM predictions\cite{hhkm} are given for
$m_t=175$~GeV, $\mh=100$~GeV, $\alpha_s(\mz)=0.12$,
and\protect\cite{eidjeg95,hhkm} $1/\bar{\alpha}(\mmz)=128.75$.
See section 4 for the definition of $\bar{\alpha}(\mmz )$.
Heavy flavor results are obtained by combining data from
LEP and SLC\protect\cite{lephf9502}.
}
\def\afb{A_{\rm FB}}
\def\mzdata    { $ 91.1884\pm 0.0022$}
\def\gammazdata{ $ 2.4963 \pm 0.0032$}
\def\sigmahdata{ $ 41.488 \pm 0.078$ }
\def\rldata    { $ 20.788 \pm 0.032$ }
\def\afbldata  { $ 0.0172 \pm 0.0012$}
\def\ataudata  { $ 0.1418 \pm 0.0075$}
\def\aedata    { $ 0.1390 \pm 0.0089$}
\def\rbdata    { $ 0.2219 \pm 0.0017$}
\def\rcdata    { $ 0.1540 \pm 0.0074$}
\def\afbbdata  { $ 0.0997 \pm 0.0031$}
\def\afbcdata  { $ 0.0729 \pm 0.0058$}
\def\slleptdata{ $ 0.2325 \pm 0.0013$}
\def\alrdata   { $ 0.1551 \pm 0.0040$}
\def\abdata    { $ 0.841  \pm 0.053 $}
\def\acdata    { $ 0.606  \pm 0.090 $}
\def\mwdata    { $ 80.26  \pm 0.16  $}
\def\mzsm    { ---          }
\def\gammazsm{ $  2.4985 $}
\def\sigmahsm{ $ 41.462  $}
\def\rlsm    { $ 20.760  $}
\def\afblsm  { $  0.0168 $}
\def\atausm  { $  0.1486 $}
\def\aesm    { $  0.1486 $}
\def\rbsm    { $  0.2157 $}
\def\rcsm    { $  0.1722 $}
\def\afbbsm  { $  0.1041 $}
\def\afbcsm  { $  0.0746 $}
\def\slleptsm{ $  0.2313 $}
\def\alrsm   { $  0.1486 $}
\def\absm    { $  0.935  $}
\def\acsm    { $  0.669  $}
\def\mwsm    { $  80.40  $}
\def\mzpull    { ---         }
\def\gammazpull{ $ -0.7 $}
\def\sigmahpull{ $  0.3 $}
\def\rlpull    { $  0.9 $}
\def\afblpull  { $  0.4 $}
\def\ataupull  { $ -0.9 $}
\def\aepull    { $ -1.1 $}
\def\rbpull    { $  3.7 $}
\def\rcpull    { $ -2.5 $}
\def\afbbpull  { $ -1.4 $}
\def\afbcpull  { $ -0.3 $}
\def\slleptpull{ $  0.9 $}
\def\alrpull   { $  1.6 $}
\def\abpull    { $ -1.8 $}
\def\acpull    { $ -0.7 $}
\def\mwpull    { $  0.9 $}
{\footnotesize
\begin{tabular}{|lr|c|c|r|}
\hline
& & data  &  SM  & ${\rm \frac{\langle data\rangle-SM}{(error)}}$
\\
\hline
{\bf LEP} & & & & \\[-2mm]
&\multicolumn{1}{l|}{line shape:} & & &\\
&${ \mz}$(GeV)       & \mzdata     & \mzsm     & \mzpull     \\[0.5mm]
&${\Gamma _Z}$(GeV)  & \gammazdata & \gammazsm & \gammazpull \\[0.5mm]
&${ \sigma_h^0}$(nb) & \sigmahdata & \sigmahsm & \sigmahpull \\[0.5mm]
&${ R_\ell\equiv}\Gamma_{ h/}\Gamma_{ \ell}$
                     & \rldata     & \rlsm     & \rlpull     \\[0.5mm]
&${ \afb^{0,\ell}}$  & \afbldata   & \afblsm   & \afblpull   \\[0.5mm]
&\multicolumn{1}{l|}{$\tau$ polarization:} & & &\\[0.5mm]
&${ A_\tau}$         & \ataudata   & \atausm   & \ataupull \\[0.5mm]
&${ A_e}$            & \aedata     & \aesm     & \aepull   \\[0.5mm]
&\multicolumn{1}{l|}{heavy flavor results:}
 & & &\\[0.5mm]
&${ R_b\equiv}\Gamma_{ b/}\Gamma_{ h}$
                     & \rbdata     & \rbsm     & \rbpull   \\[0.5mm]
&${ R_c\equiv}\Gamma_{ c/}\Gamma_{ h}$
                     & \rcdata     & \rcsm     & \rcpull   \\[0.5mm]
&${ \afb^{0,b}}$     & \afbbdata   & \afbbsm   & \afbbpull \\[0.5mm]
&${ \afb^{0,c}}$     & \afbcdata   & \afbcsm   & \afbcpull \\[1.5mm]
&\multicolumn{1}{l|}{$ q\bar{q}$ charge asymmetry:}
& & &\\[0.5mm]
&{ $ \sin^2\theta^{\rm lept}_{\rm eff}(\langle Q_{FB} \rangle)$}
                     & \slleptdata & \slleptsm & \slleptpull \\[1.5mm]
{\bf SLC} & & & & \\[-2mm]
&${ A_{LR}^0}$       & \alrdata    & \alrsm    & \alrpull \\[0.5mm]
&${ A_b}$            & \abdata     & \absm     & \abpull  \\[0.5mm]
&${ A_c}$            & \acdata     & \acsm     & \acpull  \\[1.5mm]
{\bf $ p\bar{p}$}& & & &\\[-2mm]
&${ m_W}$            & \mwdata     & \mwsm     & \mwpull  \\[1.5mm]
\hline
\end{tabular}
}
\end{center}
\end{table}
%

\begin{table}[b]
\begin{minipage}[t]{2.53in}
\tcaption{%
The error correlation matrix for the $Z$ line-shape
parameters\protect\cite{lepewwg9502}.
}
\label{tableofcorrzlineshape}
{\footnotesize
\begin{tabular}{|c|r@{\,\,}r@{\,\,}r@{$\,\,$}r@{$\,\,$}r|}
\hline
&
\multicolumn{1}{c}{$\mz$} &
\multicolumn{1}{c}{$\Gamma_Z$} &
\multicolumn{1}{c}{$\sigma_{\rm h}^0$} &
\multicolumn{1}{c}{$R_\ell$} &
\multicolumn{1}{c|}{$A_{\rm FB}^{0,\ell}$}
\\ \hline
$\mz$                 &   1.00 &$-$0.08 &   0.02 &   0.00 & 0.08\\
$\Gamma_Z$            &$-$0.08 &   1.00 &$-$0.12 &$-$0.01 & 0.00\\
$\sigma_{\rm h}^0$    &   0.02 &$-$0.12 &   1.00 &   0.15 & 0.01\\
$R_\ell$              &   0.00 &$-$0.01 &   0.15 &   1.00 & 0.00\\
$A_{\rm FB}^{0,\ell}$ &   0.08 &   0.00 &   0.01 &   0.00 & 1.00\\
\hline
\end{tabular}
}
\end{minipage}
\hfill
\begin{minipage}[t]{3.36in}
\tcaption{%
The error correlation matrix for the $b$ and $c$
quark results\protect\cite{lephf9502}.
}
\label{tableofcorrbc}
{\footnotesize
\begin{tabular}{|c|r@{\,\,}r@{\,\,}r@{\,\,}r@{\,\,}r@{\,\,}r@{\,\,}|}
\hline
&
\multicolumn{1}{c}{$R_b$} &
\multicolumn{1}{c}{$R_c$} &
\multicolumn{1}{c}{$A_{\rm FB}^{0,b}$} &
\multicolumn{1}{c}{$A_{\rm FB}^{0,c}$} &
\multicolumn{1}{c}{$A_b$} &
\multicolumn{1}{c|}{$A_c$}
\\ \hline
$R_b$             & 1.000  &$-$0.345&  0.005 &  0.055 &$-$0.068&  0.046\\
$R_c$             &$-$0.345&  1.000 &  0.084 &$-$0.063&  0.074 &$-$0.061\\
$A_{\rm FB}^{0,b}$& 0.005  &  0.084 &  1.000 &  0.109 &  0.062 &$-$0.025\\
$A_{\rm FB}^{0,c}$& 0.055  &$-$0.063&  0.109 &  1.000 &$-$0.018&  0.073\\
$A_b$             &$-$0.068&  0.074 &  0.062 &$-$0.018&  1.000 &  0.074\\
$A_c$             & 0.046  &$-$0.061&$-$0.025&  0.073 &  0.074 &  1.000\\
\hline
\end{tabular}
}
\end{minipage}
\end{table}

Before starting discussions on the implications of the new
$R_b$ and $R_c$ data in section~3, I would like you to keep
in mind the following three observations:
\begin{itemize}
\item{
The three line-shape parameters, $\Gamma_Z$, $\sigma_h^0$
and $R_l$, are now measured with accuracy better than 0.2\%.
\bsub \label{del_lineshape}
\bea
	\frac{\Delta\Gamma_Z}{\Gamma_Z} &=& -0.0010 \pm 0.0013 ,\\
	\frac{\Delta\sigma_h^0}{\sigma_h^0} &=& 0.0006 \pm 0.0019 ,\\
	\frac{\Delta R_l}{R_l} &=& 0.0015 \pm 0.0015
\eea
\esub
where $\Delta$ gives the difference between the data and the SM
predictions of Table~1.
It is important to note that these high accuracy data are sensitive
to quantum effects and that any attempted modification of the SM
should pass these tests.
}
\item{
I show in the Tables only the results obtained by assuming
the $e$-$\mu$-$\tau$ universality,
because detailed tests\cite{lepewwg9502,renton} show that a hint
of universality violation in the $\tau$-data is disappearing.
\bsub \label{e_mu_tau}
\bea
	\frac{\Delta\Gamma_\tau}{\Gamma_\tau} &=& 0.000 \pm 0.0035 ,\\
	\frac{\Delta A_\tau}{A_\tau} &=& -0.04 \pm 0.05		,\\
	\frac{\Delta A_{\rm FB}^\tau}{A_{\rm FB}^\tau} &=& 0.23 \pm 0.14 .
\eea
\esub
Although the $\tau$ Forward-Backward asymmetry is still 1.7$\sigma$
away from the SM prediction the accuracy of the measurement is still
poor and its significance is overshadowed by excellent agreements
in the partial width $\Gamma_\tau$ and the $\tau$ polarization
asymmetry which are measured at the 0.35\% and 5\% level, respectively.
}
\item{
All the asymmetry data, including the left-right beam-polarization
asymmetry, $A_{\rm LR}$, from SLC, are now consistent with each other.
I show in Fig.~1 the result of the one-parameter fit to all the
asymmetry data in terms of the effective electroweak mixing angle,
$\sbar^2(\mmz )$\cite{hhkm}, which is numerically related to the
effective parameter $\sin^2\theta_{\rm eff}^{\rm lept}$ adopted by
the LEP group\cite{lepewwg9502} as
$\sbar^2(\mmz )=\sin^2\theta_{\rm eff}^{\rm lept}-0.0010$\cite{hhkm},
within the SM.
The fit gives
\bea
\label{sb2_fit95}
	\sbar^2(\mmz ) = 0.23039 \pm 0.00029
\eea
with $\chi^2_{\rm min}/({\rm d.o.f.})=13.0/(9)$.
The updated measurements of the asymmetries agree well (16\%CL)
with the ansats that the asymmetries are determined by the
universal electroweak mixing parameter.
}
\end{itemize}
In concluding the section, I would like to point out that the
electroweak data of Tables~1--3 all together are still consistent
with the reference predictions of the SM shown in Table~1
at the 3\%CL.
The low confidence level can be traced back to the poor agreement of
the measured and predicted values for $R_b$ and $R_c$.
It does not improve significantly by varying the SM parameters
if we respect the $m_t$ bounds from the Tevatron
experiments\cite{mt_cdf,mt_d0} which typically give
160~GeV$\simlt m_t \simlt $200~GeV.

\begin{figure}[t]
\begin{center}
  \leavevmode\psfig{file=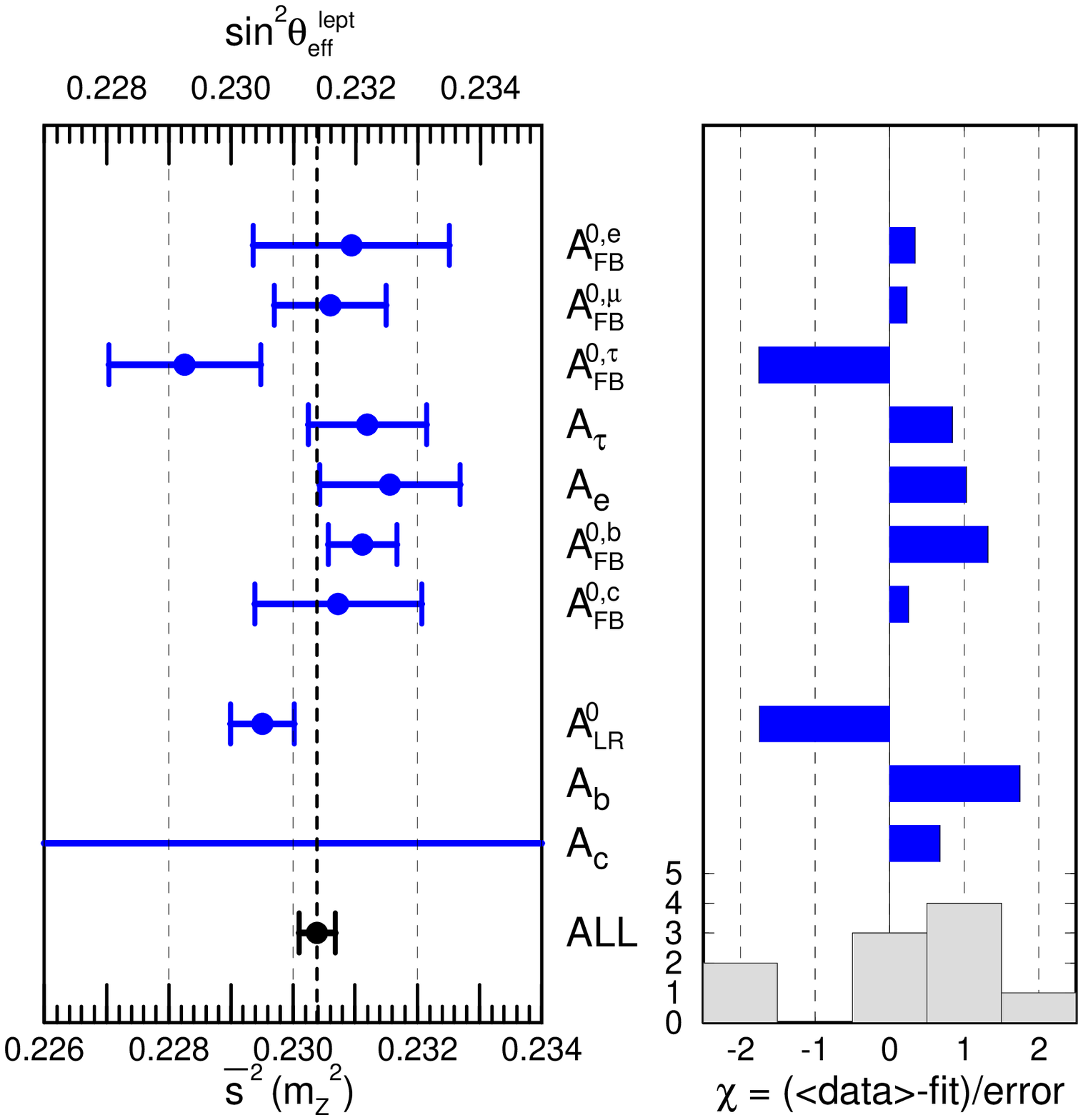,height=8cm,silent=0}
\vspace{3mm}
\fcaption{%
The effective electroweak mixing parameter
${\sbar}^2(\mmz)$ is determined from all the asymmetry
data from LEP and SLC.
The effective parameter $\sin^2\theta_{\rm eff}^{\rm lept}$
of the LEP Electroweak Working Group\protect\cite{lepewwg9502}
is related to ${\sbar}^2(\mmz)$ as\protect\cite{hhkm}
$\sin^2\theta_{\rm eff}^{\rm lept} = {\sbar}^2(\mmz) +0.0010$.
The data on $A_b$ is off the scale.
}
\label{fig:sb2}
\end{center}
\end{figure}

\section{ The $R_b$ and $R_c$ Crisis and $\alpha_s$ }

The most striking results of the updated electroweak data are
those of $R_b$ and $R_c$, which are shown in Fig.~2.
The SM predictions to these ratios are shown by the thick
solid line, where the top-quark mass in the $Zb_Lb_L$ vertex
correction is indicated by solid blobs.
When combined the $R_b$ and $R_c$ data alone reject the SM
at the 99.99\%CL.
The thin solid line represents the prediction of the extended
SM where in the $Zb_Lb_L$ vertex function\cite{hhkm},
\bea \label{zblbl}
\Gamma_L^{Zbb}(q^2) = -\gzhat \{ -\frac{1}{2}[1+\delb(q^2)]
		+\frac{1}{3}\shat^2[1+\Gamma_1^{b_L}(q^2)] \} ,
\eea
the function $\delb(\mmz )$ is allowed to take an arbitrary value.
Here $\gzhat \equiv \ghat/\chat \equiv \ehat/\shat\chat$ are properly
renormalized $\msbar$ couplings\cite{hhkm}.
In the SM, the function $\delb(\mmz )$ always takes a negative value
($\delb(\mmz )\simlt -0.03$) and its magnitude grows quadratically
with $m_t$.
It can be parametrized accurately in the region
100$<m_t(\gev )<$200 as\cite{hhkm}
\bea
\label{delb_sm}
\delb(\mmz )_{\rm SM} \approx -0.00099-0.00211(\frac{m_t+31}{100})^2 \,.
\eea
The data are not only inconsistent with the SM but also inconsistent
at more than the 2-$\sigma$ level with its extension where only
the $Zb_Lb_L$ vertex function is modified.

The correlation between the two observables, $R_b$ and $R_c$,
can be understood as follows\cite{lephf9502,renton}:
To a good approximation, the measurement of $R_c$ does not
depend on the assumed value of $R_b$, because it is measured
by detecting leading charmed-hadrons in a leading-jet for
which a b-quark jet rarely contributes.
On the other hand, the measurement of $R_b$ is affected by
the assumed value of $R_c$, since it typically makes use of
its decay-in-flight vertex signal for which charmed particles
can also contribute.
We find that the following parametrization,
\bsub \label{rb_vs_rc}
\bea
\label{rb_for_rc}
	R_b &=& 0.2205 -0.0136\frac{R_c-0.172}{0.172} \pm0.0016 ,\\
\label{rc_dat}
	R_c &=& 0.1540 \pm 0.0074 ,
\eea
\esub
reproduces the correlation obtained by the data of Tables~1 and 3
excellently, as indicated by the shaded regions in Fig.~2.
Note that we give a 39\%CL contour in all of the two-parameter
fits so that the projected 1-$\sigma$ errors can easily be
read off.

\begin{figure}[b]
\begin{minipage}[t]{7.5cm}
  \leavevmode\psfig{file=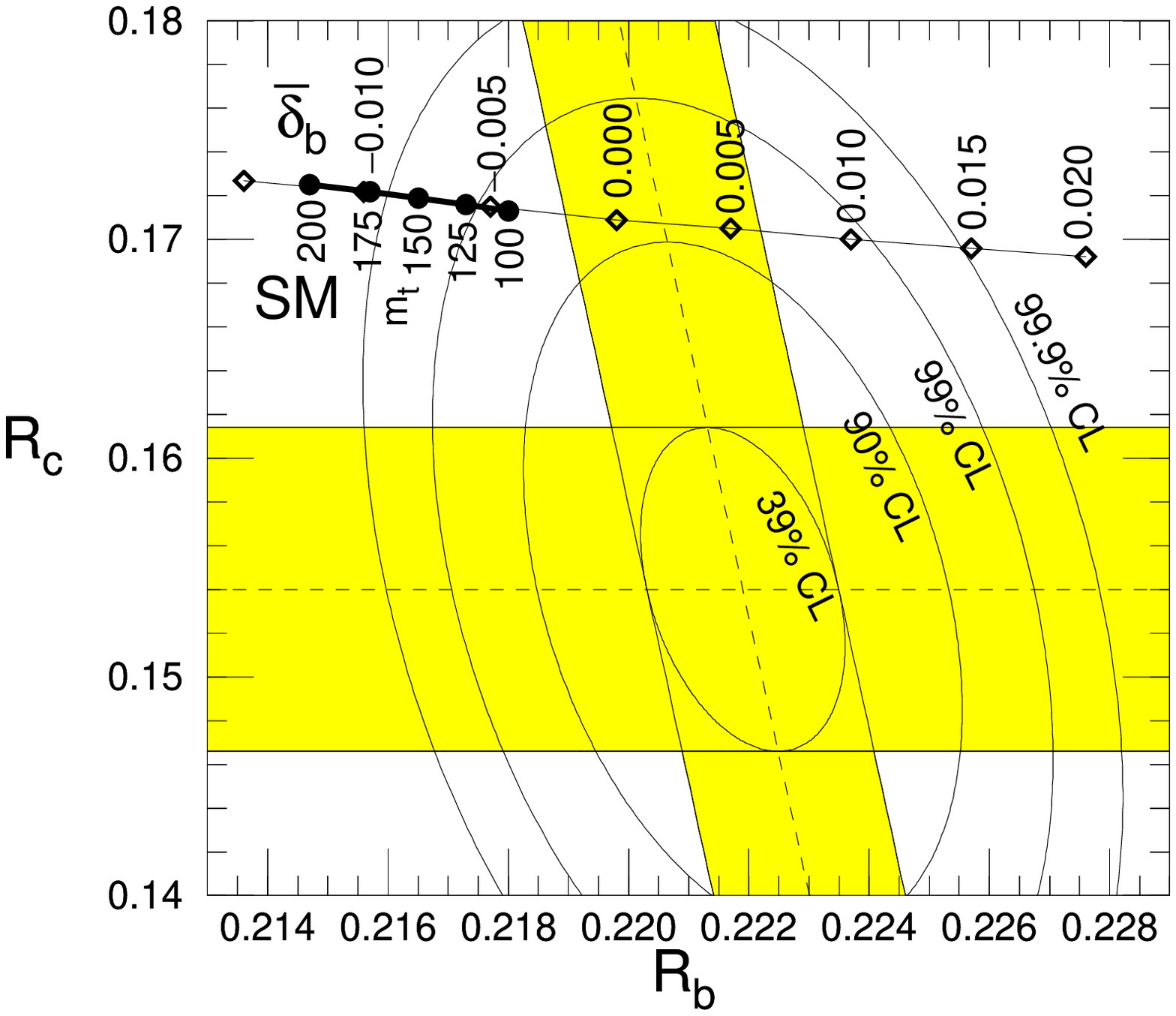,height=6cm,silent=0}
   \vspace{2mm}
\flushright{
  \begin{minipage}[t]{7.1cm}
	\fcaption{%
	$R_b$ and $R_c$ data\cite{lephf9502}
	and the SM predictions\cite{hhkm}.
	}
	\label{fig:rbrc}
  \end{minipage}
}
\end{minipage}
\hfill
\begin{minipage}[t]{7.5cm}
  \leavevmode\psfig{file=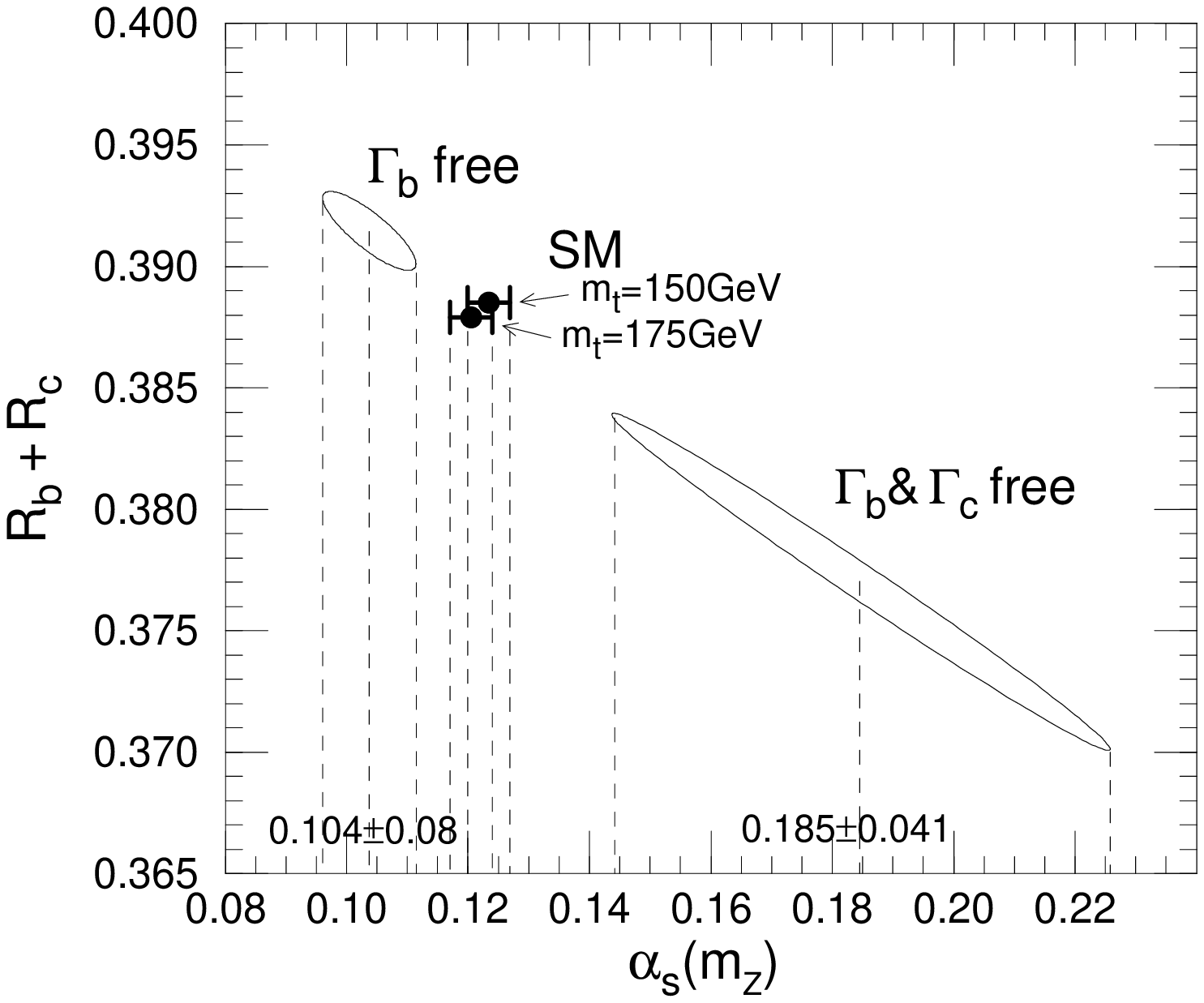,height=6cm,silent=0}
   \vspace{2mm}
\flushright{
  \begin{minipage}[t]{7.1cm}
	\fcaption{%
	$R_b+R_c$ vs $\alpha_s$.
	}
	\label{fig:rbrc_vs_alphas}
  \end{minipage}
}
\end{minipage}
\end{figure}

Before discussing the implications of this striking result,
we should remind the fact that the three most accurately
measured line-shape parameters in Eq.(\ref{del_lineshape})
determine the $Z$ partial widths $\Gamma_l$, $\Gamma_h$
and $\Gamma_{\rm inv}$ accurately,
\bsub \label{del_zwidths}
\bea
	\Delta \Gamma_h/\Gamma_h &=& \phantom{-}0.0001 \pm 0.0017 ,\\
	\Delta \Gamma_l/\Gamma_l &=& -0.0013 \pm 0.0016 ,\\
	\Delta \Gamma_{\rm inv}/\Gamma_{\rm inv} &=& -0.004 \pm 0.005 ,
\eea
\esub
because they are three independent combinations of the above
three widths,
$\Gamma_Z =\Gamma_h+3\Gamma_l+\Gamma_{\rm inv}$,
$R_l=\Gamma_h/\Gamma_l$, and
$\sigma_h^0 = (12\pi/\mmz )\Gamma_h\Gamma_l/\Gamma_Z^2$.
That the hadronic $Z$ partial width, $\Gamma_h$, is measured
with 0.17\% accuracy strongly constrains our attempt to
modify theoretical predictions for the ratios $R_b$ and $R_c$.
This is because $\Gamma_h$ can be approximately expressed as
\bea
	\Gamma_h &=& \Gamma_u +\Gamma_d +\Gamma_s +\Gamma_c +\Gamma_b
		+\Gamma_{\rm others}
\nonumber\\
	&\sim& \{ \Gamma_u^0 +\Gamma_d^0 +\Gamma_s^0
	+\Gamma_c^0 +\Gamma_b^0 \}
	\times [1+\frac{\alpha_s}{\pi}+ {\cal O}(\frac{\alpha_s}{\pi})^2 ] ,
\label{gamma_h_appr}
\eea
where $\Gamma_q^0$'s are the partial widths in the absence of the
final state QCD corrections.
Hence, to a good approximation, the ratios $R_q$ can be expressed
as ratios of $\Gamma_q^0$ and their sum.
A decrease in $R_b$ and an increase in $R_c$ should then
imply a decrease and an increase of $\Gamma_b^0$ and $\Gamma_c^0$,
respectively, from their SM predicted values.
In order to satisfy the experimental constraint on $\Gamma_h$
one should hence adjust the $\alpha_s$ value in Eq.(\ref{gamma_h_appr}).

The consequence of this constraint is clearly shown in Fig.~3
where, once we allow {\em both} $\Gamma_b^0$ {\em and} $\Gamma_c^0$
to be freely fitted by the data, the above $\Gamma_h$
constraint forces $\alpha_s$ to be unacceptably large.
On the other hand, if we allow only $\Gamma_b^0$ to vary
by assuming the SM value of $\Gamma_c^0$
(the straight line of the extended SM in Fig.~2),
then the $\Gamma_h$ constraint gives a slightly small value
of $\alpha_s$, which is compatible\cite{hhkm,shifman}
with some of the low-energy measurements\cite{pdg94,ccfr95}
and lattice QCD estimates\cite{elkhadra92,davies95}.
Although the SM does not reproduce the $R_b$ and $R_c$ data
it gives a moderate $\alpha_s$ value consistent with the
estimates based on the $e^+e^-$ jet-shape
measurements\cite{bethke95,sld95}
and the hadronic $\tau$-decay rate\cite{bethke95}.
Although the $\tau$-decay measurement has the smallest
experimental and perturbative-QCD error it may still
suffer from non-perturbative corrections\cite{truong93},
and a larger theoretical uncertainty may be
assigned\cite{altarelli95}:  see Fig.~4.

\begin{figure}[t]
\begin{center}
  \leavevmode\psfig{file=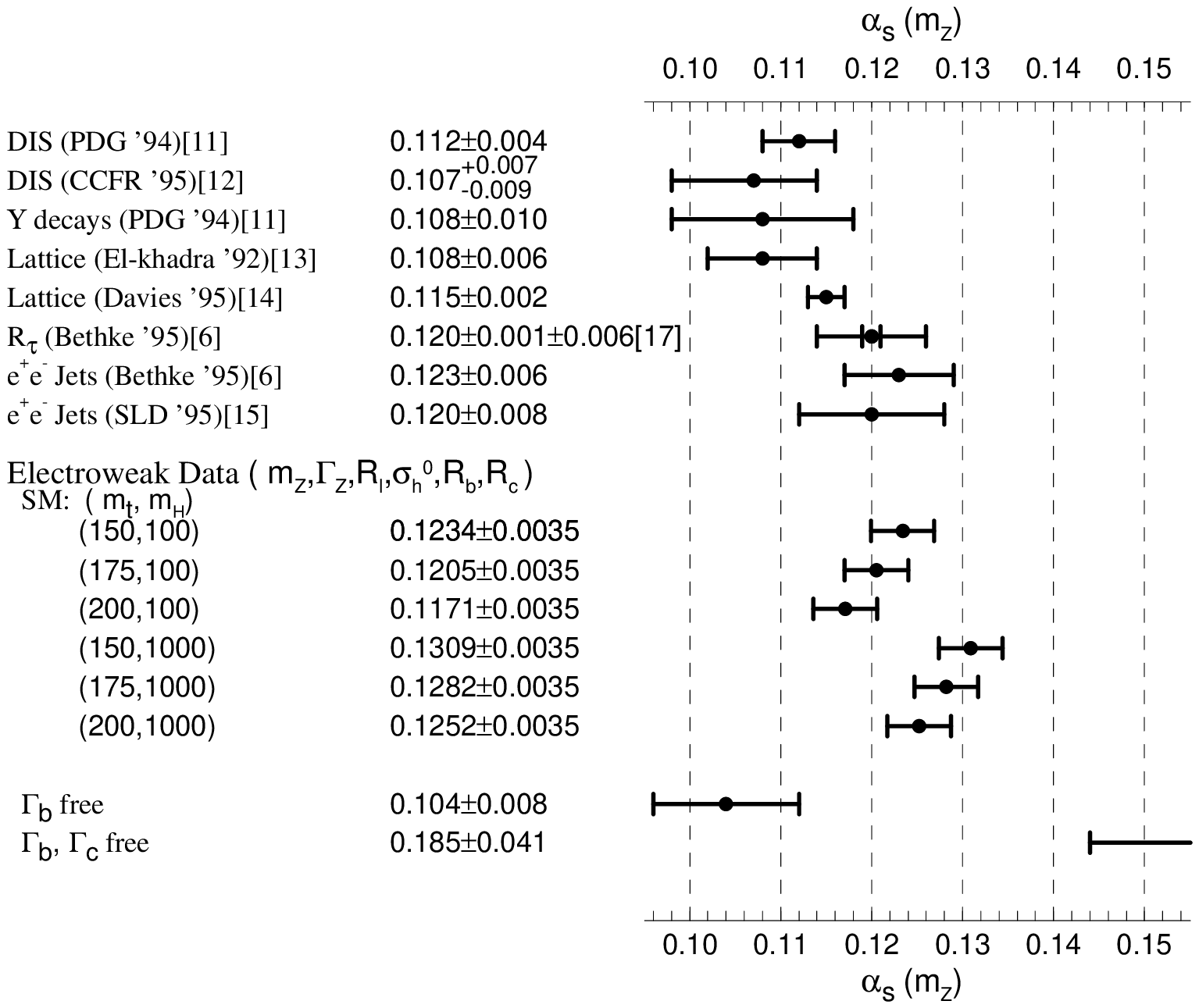,height=8cm,silent=0}
\vspace{2mm}
\fcaption{%
Compilation of $\alpha_s$ values as measured by various
experiments and by Lattice QCD calculations.
}
\end{center}
\end{figure}

In fact we do not yet have a definite clue where
in the region $0.105<\alpha_s(\mz )<0.125$ the true
QCD coupling constant lies.
$\alpha_s\sim 0.12$ is favored from the electroweak
data, if we believe in the SM predictions for $R_b$ and
$R_c$ despite the strong experimental signals.
On the other hand $\alpha_s\sim 0.11$ is favored
if we believe in the SM predictions for $\Gamma_c^0$
while allowing new physics to modify $\Gamma_b^0$.
These two solutions both lead to an acceptable
$\alpha_s$ value at present.
However, once we allow new physics in both $\Gamma_b^0$
and $\Gamma_c^0$ and let them be fitted by the data,
then an unacceptably large $\alpha_s$ follows.

In fact, if we stick to the very conservative bounds
$0.105<\alpha_s(\mz )<0.125$, we cannot explain the
discrepancies in both $R_b$ and $R_c$ by allowing
new physics only in $\Gamma_b^0$ and $\Gamma_c^0$.
The only sensible solution, then, may be to allow
a new physics contribution to all $\Gamma_q^0$
such that their sum stays roughly at the SM value,
e.g. by making all the down-type-quark widths larger
than their SM values by 3\% and the up-type-quark
widths to be smaller by 6\%.
Such a model would explain both $R_b$ and $R_c$,
and give a reasonable $\alpha_s$.
It is not easy to find a working model, however, which does
not jeopardize all the excellent successes of the SM
in the quark and lepton asymmetries, the leptonic widths,
$\mw$, and in the low-energy neutral-current data.

Because the $R_c$ measurement depends strongly on the
charm-quark detection efficiency,\, which has uncertainties
in charmed-quark fragmentation function into charmed
hadrons and in charmed-hadron decay branching fractions,
it is still possible that unexpected errors are hiding.
As an extreme example, if as much as 10\% of the charmed-quark
final states were unaccounted for, then {\em both} the
10\% deficit in $R_c$ at LEP {\em and} the 10\% too few
charmed hadron multiplicity in B-meson
decays\cite{buchalla95,neubert95}
can be solved.

If we assume that $R_c$ actually has the SM value $R_c\sim 0.172$
and temporarily set aside its experimental constraint,
then the correlation as depicted by Eq.(\ref{rb_for_rc}) tells
that the measured $R_b=0.2205\pm0.0016$ is about 2\% larger
than the SM prediction, $R_b\sim 0.216$ for $m_t\sim 175$~GeV.
The discrepancy is still significant at the 3-$\sigma$ level.

I examined the possibility that an experimental problem
that could result in an underestimation of $R_c$ can lead to
an overestimation of $R_b$.
This does not seem to be the case, since $R_b$ is measured
mainly by using a different technique, the double-tagging
method\cite{renton}, where the $b$-quark tagging efficiency is
determined experimentally rather than by estimating it from
the $b$-quark fragmentation model and the $b$-flavored
hadron decay rates.  Schematically the single and double
$b$-tag event rate in hadronic two-jet events are expressed as
\bsub \label{b_tagg}
\bea
\frac{N_T}{2N_h} &=&  	 \epsilon_b r_b R_b
			+\epsilon_c r_c R_c
			+\mbox{\rm others} ,
\\
\frac{N_{TT}}{N_h} &=&  C\,\{\,\epsilon_b^2 r_b R_b
			+\epsilon_c^2 r_c R_c
			+\mbox{\rm others} \,\} ,
\eea
\esub
where $\epsilon_q$ denotes the efficiency of tagging a $q$-jet,
$r_q$ is the rate of two-jet-like events ($T>0.8$) in the
$q\bar{q}$ initiated events, and deviation of $C$ from unity
measures possible correlation effects between two jets.
By choosing the tagging condition such that
$\epsilon_b \gg \epsilon_c$, one can self-consistently
determine both $\epsilon_b$ and $R_b$:
\bea
\label{rb_btagg}
R_b\, = \,\frac{C}{r_b}\,
\frac{[\,\frac{N_T}{2N_h} -\epsilon_c\frac{r_c}{r_b}R_c -\cdot\cdot\cdot\,]^2}
   {[\,\frac{N_{TT}}{N_h} -\epsilon_c^2\frac{r_c}{r_b}R_c -\cdot\cdot\cdot\,]}
\,.
\eea
In the limit of uncorrelated two-jet events only $C=1$ and $r_b=1$,
and in the limit of a negligible contribution from non-$b$ events,
the ratio $R_b$ is determined from the ratio of the square of
the single-tag event rate and the double-tag event rate.
Only for the corrections to this limit are the QCD motivated
hadron-jet Monte Carlo programs used.
A compilation of very careful tests of these correction terms are
found in the LEP/SLC Heavy Flavor Group report\cite{lephf9502}.
We should still examine if our present understanding of generating
hadronic final states from quark-gluon states allows us to constrain
the coefficients $C$ and $r_b$ at much less than a \% level
and the miss-tagging efficiency $\epsilon_c\sim 0.01$ at 10\% level.
For instance, the combination of an overestimation of $C$ by 0.5\%
with an underestimation of $r_b$ by 0.5\% and $\epsilon_c$ by 10\%,
can result in an overestimate of $R_b$ by 2\%.
Serious theoretical studies of the uncertainty in the present
hadron-jet generation program are needed, because in my opinion,
these programs have never been tested at the accuracy level that
was achieved by these excellent experiments at LEP.

There have been many attempts to explain the discrepancy in
$R_b$ by invoking new physics beyond the SM.
Most notably, in the minimal supersymmetric (SUSY) SM\cite{boul,abc,%
wells1,kimpark,sola1,rc,dabel,wells2,chankow2,wagner,wang,ma,wells3,yhm95},
an additional loop of a light $\tilde{t}_R$ and a light
higgsino-like chargino, or that with an additional Higgs pseudoscalar
when $\tan\beta\gg 1$, can compensate the large negative
top quark contribution of the SM in the $Zb_Lb_L$ vertex function.
Such a solution typically leads to the prediction that the masses
of the lighter $\tilde{t}$ and chargino, or the pseudoscalar
should be smaller than $\mz$.
In the former scenario the top quark should have significant exotic
decays into $\tilde{t}_R$ and a neutral Higgsino, and in the latter
scenario another exotic decay $t\to b+H^+$ may
occur\cite{wells2,wang,ma,wells3,yhm95}.
In both SUSY scenarios, we should expect to find new particles at
Tevatron, LEP2 or even at LEP1.5.

\begin{figure}[t]
\begin{center}
  \leavevmode\psfig{file=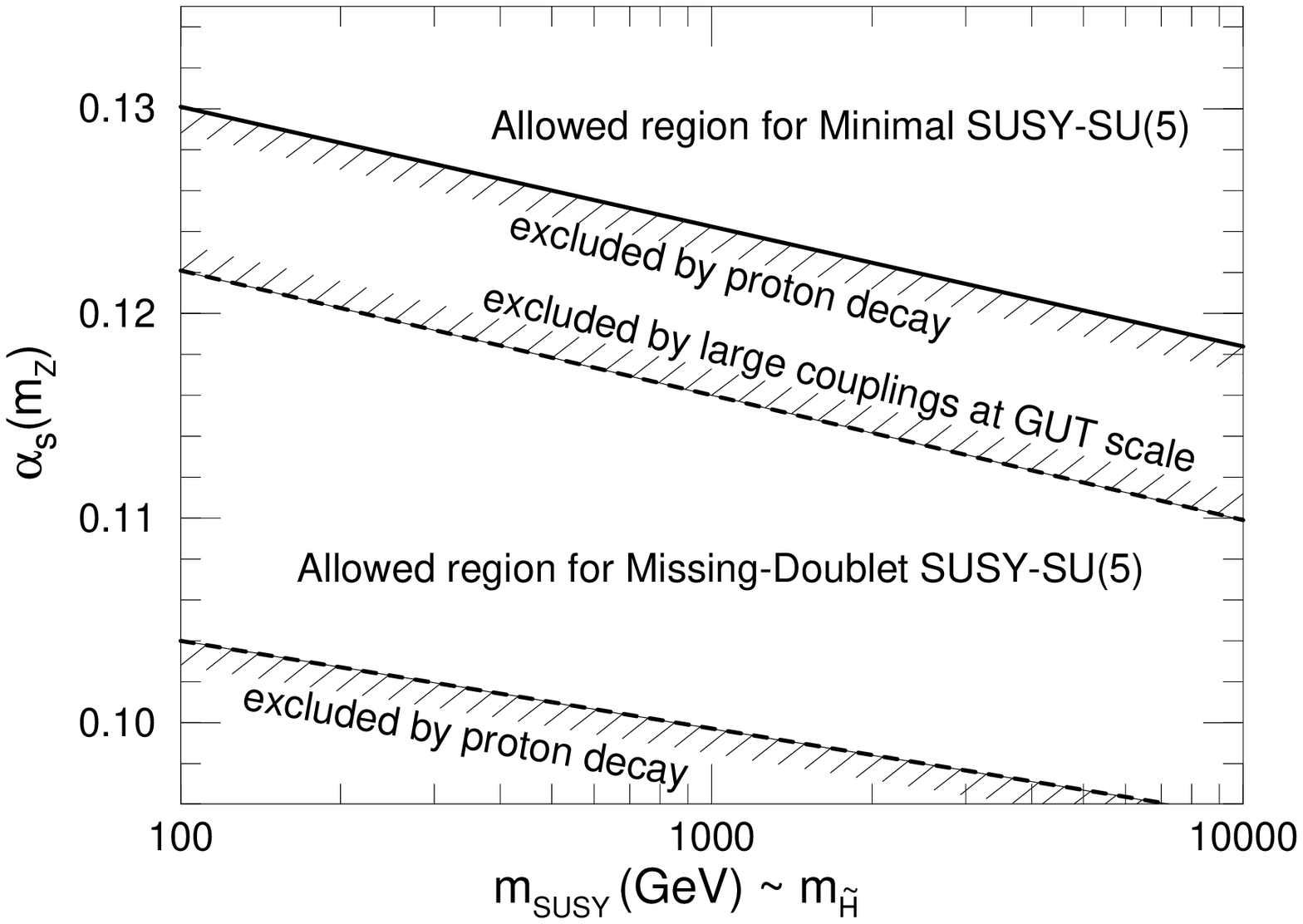,height=7cm,silent=0}
\fcaption{%
Constraints on $\alpha_s(\mz)_{\msbar}$ as functions of the
SUSY threshold scale $m_{\rm SUSY}$ in the minimal SUSY-SU(5)
model\cite{minsu5} and in the model with the missing-doublet
mechanism\cite{mdmsu5}.
}
\end{center}
\end{figure}

In an alternative scenario of the electroweak symmetry breaking,
the Techni-Color (TC) model, the heavy top quark mass implies
strong interactions among top-quarks and techniquarks.
Such interactions, typically called the extended technicolor
(ETC) interactions, can affect the $Zb_Lb_L$ vertex.
However, the side-ways ETC bosons that connect the top-quark and
techniquark leads to a contribution with an opposite
sign\cite{chivukula92,evans94}.
The diagonal ETC bosons contribute\cite{kitazawa93} with
the correct sign\cite{ghwu95}, and their consequences have
been studied\cite{yue95,hk95}.
The diagonal ETC bosons that explain the $R_b$ data are,
however, found\cite{yoshikawa95} to give an unacceptably large
contribution to the $T$ parameter\cite{stu}.

As an alternative to the standard ETC model where the ETC
gauge group commutes with the SM gauge group, models with
non-commuting ETC gauge group have been proposed\cite{chivukula94}
which have rich phenomenological consequences.
It is also noted that an existence of the new heavy gauge boson
$X$ that couples only to the third-generation quarks and leptons
has been proposed\cite{holdom94}, which can affect both the
$Zbb$ and $Z\tau\tau$ vertices through mixing with the SM $Z$.

So far, the above models affect mainly the $Zb_Lb_L$ coupling,
which dominates the $Zb_Rb_R$ coupling in the SM.
A possible anomaly in the $b$-jet asymmetry parameter, $A_b$,
observed at SLC with its polarized beam, see Table~1 and Fig.~1,
may suggest a new physics contribution in the $Zb_Rb_R$
vertex\cite{peskin95}.
It is worth watching improved $A_b$ measurements at SLC
in the future.

Finally, I would like to note that the small $\alpha_s$ value
which is obtained by allowing a new physics contribution to explain
the $R_b$ anomaly tends to destroy the SUSY-SU(5) unification
of the three gauge couplings in the minimal model\cite{minsu5}.
This problem is, however, highly dependent on details of
the particle mass spectrum at the GUT scale.
In fact in the missing doublet SUSY-SU(5) model\cite{mdmsu5}
which naturally explains the doublet-triplet splitting,
smaller $\alpha_s$ is prefered due to its peculiar
GUT particle spectrum\cite{hy92,bagger95,lopez95,jellis95}.
I show in Fig.~5 the update\cite{hmy96} for the allowed regions
of $\alpha_s(\mz )$ in the two SUSY-SU(5) models as functions
of the heavy Higgsino mass, where the standard supergravity model
assumptions are made for the SUSY particle masses at the
electroweak scale.

\section{ Global Fit to All Electroweak Data with $S$, $T$, $U$ }

In this section we present the results of the global fit to all
the electroweak data in which we allow new physics contribution
to the $S$, $T$, $U$ parameters\cite{stu} of the electroweak
gauge-boson-propagator corrections as well as to the $Zb_Lb_L$
vertex form factor, $\delb(\mmz )$, but otherwise we assume the
SM contribution to dominate the corrections.
We take the strengths of the QCD and QED couplings at the $\mz$
scale, $\alpha_s(\mz )$ and $\bar{\alpha}(\mmz )$, as external
parameters of the fits, so that implications of their
precise measurements on electroweak physics are manifestly shown.

\subsection{ Brief Review of Electroweak Radiative Corrections
	     in $\rm{SU(2)_L \times U(1)_Y}$ Models }

The propagator corrections in the general
$\rm{SU(2)_L \times U(1)_Y}$
models can conveniently be expressed in terms of the following
four effective charge form-factors\cite{hhkm}:
\vspace*{-12mm}
\def\propagator#1#2{%
    \begin{picture}(80,35)(0,18)
       \Text(20,27)[cb]{#1}
       \Text(50,27)[cb]{#2}
       \Line(0,35)(10,20)
       \Line(0, 5)(10,20)
       \Line(70,35)(60,20)
       \Line(70, 5)(60,20)
       \Photon(10,20)(30,20){3}{3}
       \Photon(40,20)(60,20){3}{3}
       \GCirc(35,20){7}{0.5}
    \end{picture}
}
\bsub \label{barcharges}
  \begin{eqnarray}
    \propagator{$\gamma$}{$\gamma$} &\sim& \bar{\alpha}(q^2) =\ehat^2
    \Bigl[\,1-{\rm Re}\pibar_{T,\gamma}^{\gamma\gamma}(q^2)\,\Bigr]\,,
\\[-2mm]
    \propagator{$\gamma$}{$Z$}      &\sim& \sbar^2(q^2)      =\shat^2
    \Bigl[\,1+\shat\chat\,{\rm Re}\,\pibar_{T,\gamma}^{\gamma Z}(q^2)
    \,\Bigr]\,
\\[-2mm]
    \propagator{$Z$}{$Z$}           &\sim& \gzbar^2(q^2)     =\gzhat^2
    \Bigl[\,1 -{\rm Re}\pibar_{T,Z}^{ZZ}(q^2)\,\Bigr]\,,
\\[-2mm]
    \propagator{$W$}{$W$}           &\sim& \gwbar^2(q^2)     =\ghat^2
    \Bigl[\,1 -{\rm Re}\pibar_{T,W}^{WW}(q^2)\,\Bigr]\,,
  \end{eqnarray}
\esub
where
$
     \pibar_{T,V}^{AB}(q^2)
      \equiv [\pibar_T^{AB}(q^2)\!-\!\pibar_T^{AB}(\mmv)]/(q^2\!-\!\mmv)
$
are the propagator correction factors that appear in the
$S$-matrix elements after the mass renormalization is performed,
and $\ehat \equiv \ghat\shat \equiv \gzhat^{}\shat\chat$
are the $\msbar$ couplings.
The `overlines' denote the inclusion of the pinch
terms\cite{pinch,kl89}, which make these effective charges
useful\cite{kl89,hhkm,hms95} even at very high energies
($|q^2|\gg \mmz$).
The amplitudes are then expressed in terms of these charge
form-factors plus appropriate vertex and box corrections.
Hence the charge form-factors
can be directly extracted from the experimental data
by assuming SM dominance to the vertex and box corrections,
and the extracted values can be compared with various
theoretical predictions.

We can {\rm define}\cite{hhkm} the $S$, $T$, and $U$ variables
of Ref.\cite{stu} in terms these effective charges,
  \begin{subequations}\label{stu_def}
  \begin{eqnarray}
    \frac{\sbar^2(\mmz)\cbar^2(\mmz)}{\bar{\alpha}(\mmz)}
    -\frac{4\,\pi}{\gzbar^2(0)} &\equiv& \;\frac{S}{4} \,,
    \\
    \frac{\sbar^2(\mmz)}{\bar{\alpha}(\mmz)}\quad
    -\frac{4\,\pi}{\gwbar^2(0)} &\equiv& \frac{S+U}{4} \,,
    \\
    1\;-\,\frac{\gwbar^2(0)}{\mmw}\frac{\mmz}{\gzbar^2(0)}
    &\equiv& \;\alpha T \,,
  \end{eqnarray}
  \end{subequations}
where it is made clear that these variables measure deviations
from the naive universality of the electroweak gauge boson
couplings.
They receive contributions from both the SM radiative effects
as well as new physics contributions.
The original $S$, $T$, $U$ variables\cite{stu} are
obtained\cite{hhkm} approximately by subtracting
the SM contributions (at $\mh=1000$~GeV).

For a given electroweak model we can calculate the $S$, $T$,
$U$ parameters ($T$ is a free parameter in models without
the custodial SU(2) symmetry), and the charge form-factors
are then fixed by the following identities\cite{hhkm}:
  \begin{subequations}
    \label{gbarfromstu}
  \begin{eqnarray}
     \frac{1}{\gzbar^2(0)}
        &=& \frac{1+\delg -\alpha \,T}{4\,\sqrt{2}\,G_F\,\mmz} \,,
    \label{gzbarfromt}\\[2mm]
      \sbar^2(\mmz)
          &=& \frac{1}{2}
              -\sqrt{\frac{1}{4} -\bar{\alpha}^2(\mmz)
                    \biggl(\frac{4\,\pi}{\gzbar^2(0)} +\frac{S}{4}
                    \biggr)  }\,,
    \label{sbarfroms}\\[2mm]
       \frac{4\,\pi}{\gwbar^2(0)}
             &=& \frac{\sbar^2(\mmz)}{\bar{\alpha}^2(\mmz)}
                -\frac{1}{4}\,(S+U) \,.
    \label{gwbarfromu}
  \end{eqnarray}
  \end{subequations}
Here $\delg$ is the vertex and box correction to the muon
lifetime\cite{del_gf} after subtraction of the pinch term\cite{hhkm}:
 \begin{eqnarray}
   G_F &=& \frac{\gwbar^2(0)+\ghat^2\delg}{4\,\sqrt{2}\mmw} \,.
   \label{gf}
 \end{eqnarray}
In the SM, $\delg=0.0055$\cite{hhkm}.

It is clear from the above identities that once we know
$T$ and $\delg$ in a given model we can predict
$\gzbar^2(0)$, and then by knowing $S$ and $\bar{\alpha}(\mmz)$
we can calculate $\sbar^2(\mmz)$, and finally by knowing
$U$ we can calculate $\gwbar^2(0)$.
Since $\bar{\alpha}(0)=\alpha$ is known precisely, all
four charge form factors are fixed at one $q^2$ point.
The $q^2$-dependence of the form factors should also be
calculated in a given model, but it is less dependent
on physics at very high energies\cite{hhkm}.
In the following analysis we assume that the SM contribution
governs the running of the charge form-factors between
$q^2=0$ and $q^2=\mmz$.
We can now predict all the neutral-current amplitudes
in terms of $S$ and $T$, and an additional knowledge of
$U$ gives the $W$ mass via Eq.(\ref{gf}).

We should note here that our prediction for the effective
mixing parameter $\sbar^2(\mmz)$ is not only sensitive to
the $S$ and $T$ parameters but also on the precise value
of $\bar{\alpha}(\mmz)$.
This is the reason why our predictions for the asymmetries
measured at LEP/SLC and, consequently, the experimental
constraint on $S$ extracted from the asymmetry data
are dependent on $\bar{\alpha}(\mmz)$.
In order to parametrize the uncertainty in our evaluation
of $\bar{\alpha}(\mmz)$, the parameter $\delta_\alpha$
is introduced in Ref.\cite{hhkm} as follows:
$1/\bar{\alpha}(\mmz)\!\equiv\! 4\pi/\ebar(\mmz)\!
=\!128.72\!+\!\delta_\alpha$.
We show in Table~4 the results of the four recent
updates\cite{mz94,swartz95,eidjeg95,bp95}
on the hadronic contribution to the running of the effective
QED coupling.
Three definitions of the running QED coupling are compared.
I remark that our simple formulae
(\ref{barcharges}) and (\ref{stu_def}) are valid only if one
includes all the fermionic and bosonic contributions
to the propagator corrections.

\begin{table}[t]
\tcaption{%
The running QED coupling at the $\mz$ scale in the three schemes.
$1/\alpha(\mmz)_{\rm l.f.}$ contains only the light fermion
contributions to the running of the QED coupling constant
between $q^2=0$ and $q^2=\mmz$.
$1/\alpha(\mmz)_{\rm f}$ contains all fermion contributions
including the top-quark.
$m_t=175$~GeV and $\alpha_s(\mz)=0.12$ in the perturbative
two-loop correction\cite{kniehl90} are assumed.
$1/\bar{\alpha}(\mmz)$ contains also the $W$-boson-loop
contribution\cite{hhkm} including the pinch term\cite{pinch,pinch2}.
}
\label{tab:alpha_mz}
\begin{center}
{\footnotesize
\begin{tabular}{|c|c|c|c|c|}
\hline
& $1/\alpha(\mmz)_{\rm l.f.}$
& $1/\alpha(\mmz)_{\rm f}$
& $1/\bar{\alpha}(\mmz)$
& $\delta_\alpha$
\\
\hline
Martin-Zeppenfeld '94\cite{mz94}
& $128.98\pm 0.06$ & $128.99\pm 0.06$ & $128.84\pm 0.06$ & $0.12\pm 0.06$
\\
Swartz '95\cite{swartz95}
& $128.96\pm 0.06$ & $128.97\pm 0.06$ & $128.82\pm 0.06$ & $0.10\pm 0.06$
\\
Eidelman-Jegerlehner '95\cite{eidjeg95}
& $128.89\pm 0.09$ & $128.90\pm 0.09$ & $128.75\pm 0.09$ & $0.03\pm 0.09$
\\
Burkhardt-Pietrzyk '95\cite{bp95}
& $128.89\pm 0.10$ & $128.90\pm 0.10$ & $128.76\pm 0.10$ & $0.04\pm 0.10$
\\
\hline
\end{tabular}
}
\end{center}
\end{table}

A more extensive list of the estimates are shown in Fig.~6.
The analysis of Ref.\cite{hhkm} was based on the estimate
\cite{jeg92}, $\delta_\alpha=0.00\pm 0.10$.
Nevzorov \etal\cite{nevzorov94} made use of perturbative
QCD down to $\sqrt{s}\sim 1$~GeV, while the estimates by
Geshkenbein and Morgunov\cite{gm94,gm95} are based on
a theoretical model of resonance production.
Martin and Zeppenfeld\cite{mz94} also relied on the perturbative
QCD, but they restrict its use to constraining only the overall
normalization of the data at $\sqrt{s}>3$~GeV.
Their estimate agrees well with the revised evaluation by
Swartz\cite{swartz95} who used only experimental data.
Finally two of the most recent values\cite{eidjeg95,bp95}
agree perfectly.
There is no real discrepancy among the four recent estimates
in Table~4, where small differences are attributed
to the use of perturbative QCD for constraining the magnitude
of medium energy data\cite{mz94} or to a slightly different set
of input data\cite{swartz95}.
For more detailed discussions I refer the readers to
an excellent review by Takeuchi\cite{takeuchi95}.
In the following analysis we take the estimate of
Ref.\cite{eidjeg95} ($\delta_\alpha=0.03\pm 0.09$)
as the standard, and show sensitivity of our results
on $\delta_\alpha-0.03$.

\begin{figure}[b]
 \begin{center}
 \leavevmode\psfig{file=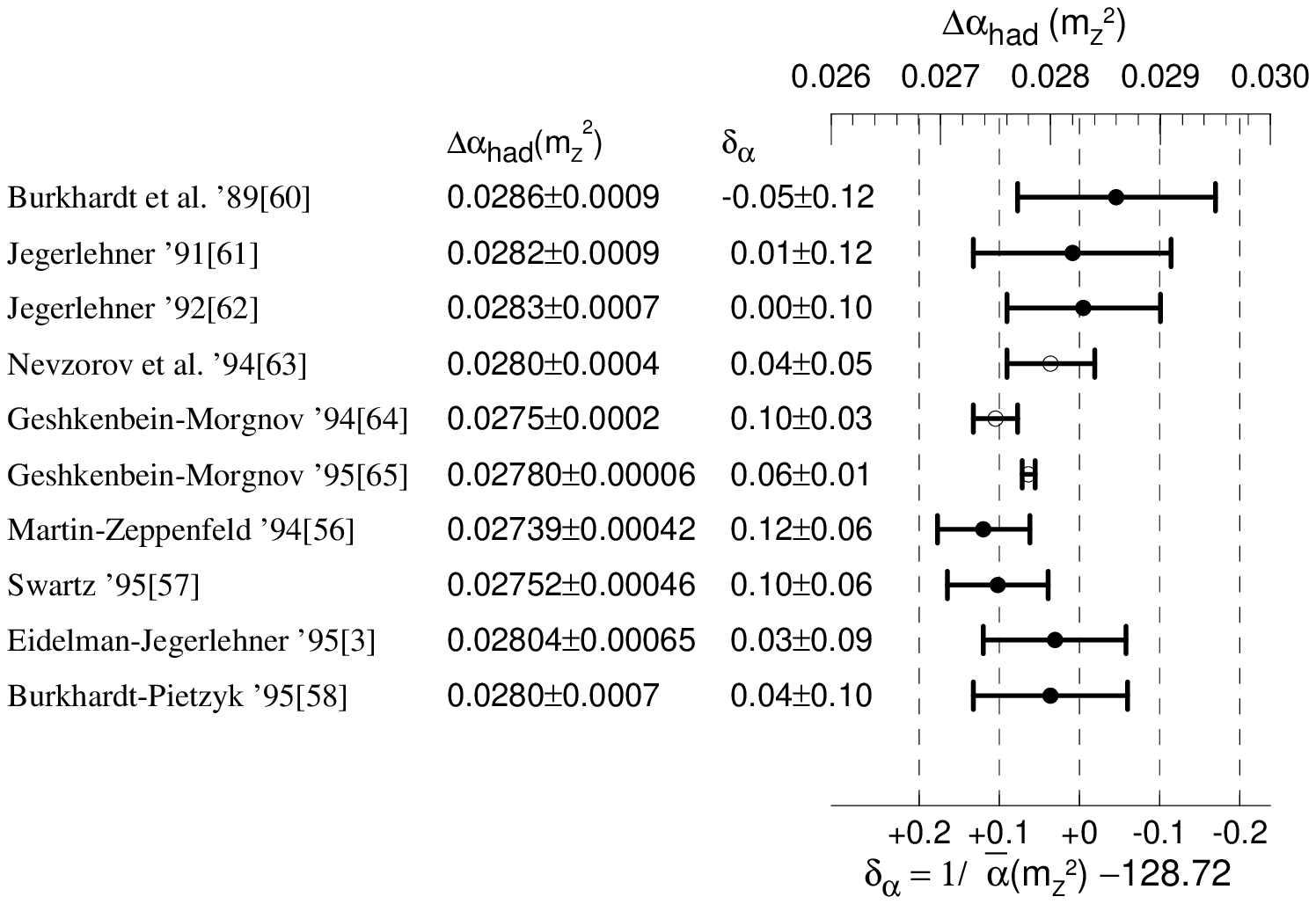,height=8cm,silent=0}
 \end{center}
\fcaption{%
Various estimates of $\Delta\alpha_{\rm had}(\mmz)$
and the resulting $\bar{\alpha}(\mmz)$ in the
minimal SM.
The parameter $\delta_\alpha$\cite{hhkm} is defined as
$\delta_\alpha \equiv 1/\bar{\alpha}(\mmz)-128.72$.
}
\label{fig:delta_had}
\end{figure}

Once we know $\bar{\alpha}(\mmz)$ the charge form-factors
in Eq.(\ref{gbarfromstu}) can be calculated from $S$, $T$, $U$.
The following approximate formulae\cite{hhkm} are useful:
  \begin{subequations}
   \label{gbar_approx}
  \begin{eqnarray}
        \gzbar^2(0)    &\approx& 0.5456 \hphantom{+0.0036\,S}\;\,
            +0.0040\,T'\,,
   \label{gzbar_approx}\\
        \sbar^2(\mmz) &\approx& 0.2334            +0.0036\,S
            -0.0024\,T'
           \hphantom{ +0.0035\,U }\;\, -0.0026\,\delta_\alpha    \,,
         \qquad
   \label{sbar_approx}\\
        \gwbar^2(0)    &\approx& 0.4183            -0.0030\,S
            +0.0044\,T'
            +0.0035\,U               +0.0014\,\delta_\alpha      \,,
         \qquad
   \label{gwbar_approx}
  \end{eqnarray}
  \end{subequations}
where $T'= T+(0.0055-\delg)/\alpha$.
The values of $\gzbar^2(\mmz)$ and $\sbar^2(0)$ are then
calculated from $\gzbar^2(0)$ and $\sbar^2(\mmz)$ above,
respectively, by assuming the SM running of the form-factors.
The $Z$ widths are sensitive to $\gzbar^2(\mmz)$, which can
be obtained from $\gzbar^2(0)$ in the SM approximately by
\gzbarrunning
when $m_t(\gev )>150$ and $\mh(\gev )>100$.
Details of the following analysis will be reported elsewhere\cite{hhm96}.

\subsection{ Global Fit to All the Electroweak Data }

By assuming that all the vertex corrections except for the
$Zb_Lb_L$ vertex function $\delb(\mmz )$ are dominated by
the SM contributions, we make a four-parameter fit to the
LEP/SLC data\footnote{%
We exclude from the fit the jet-charge asymmetry data in
Table~1, since it allows an interpretation only within the
minimal SM.  It is included in our SM fit in section 5. }
{}~of Tables~1--3 in terms of
$\gzbar^2(\mmz )$, $\sbar^2(\mmz )$, $\delb(\mmz )$ and
$\alpha_s=\alpha_s(\mz )_{\msbar}$.
We find\cite{hhm96}
%
\fitofgzbsb
%
where
\bea
\label{alps'}
\alpha_s' = \alpha_s(\mz )_{\msbar}\, +1.54\,\delb(\mmz )
\eea
is the combination\cite{hhkm} that appears in the theoretical
prediction for $\Gamma_h$.
The best fit is obtained at $\delb(\mmz )=0.0025$ and
$\alpha_s=0.1043$ as a consequence of the $R_b$ data:
see Figs.~3 and 4.
On the other hand, if we assume the SM value for $\delb(\mmz )$,
Eq.(\ref{delb_sm}), the above fit gives
$\alpha_s=0.1234\pm0.0043$ for $m_t=175$~GeV.
In Fig.~7, we show the 1-$\sigma$ (39\%CL) allowed contours
for $\alpha_s=0.115$, 0.120, 0.125 when $\delb$ takes its
SM value at $m_t=175$~GeV.
If we allow {\em both} $\delb(\mmz )$ and $\alpha_s$ to be
freely fitted by the data, we obtain the solid contour in
Fig.~7.
The SM predictions for $\delta_\alpha=0.03$ and their
dependence on $\delta_\alpha-0.03$ are also given.
As expected, only $\sbar^2(\mmz )$ is sensitive to
$\delta_\alpha$.

\begin{figure}[t]
 \begin{center}
 \leavevmode\psfig{file=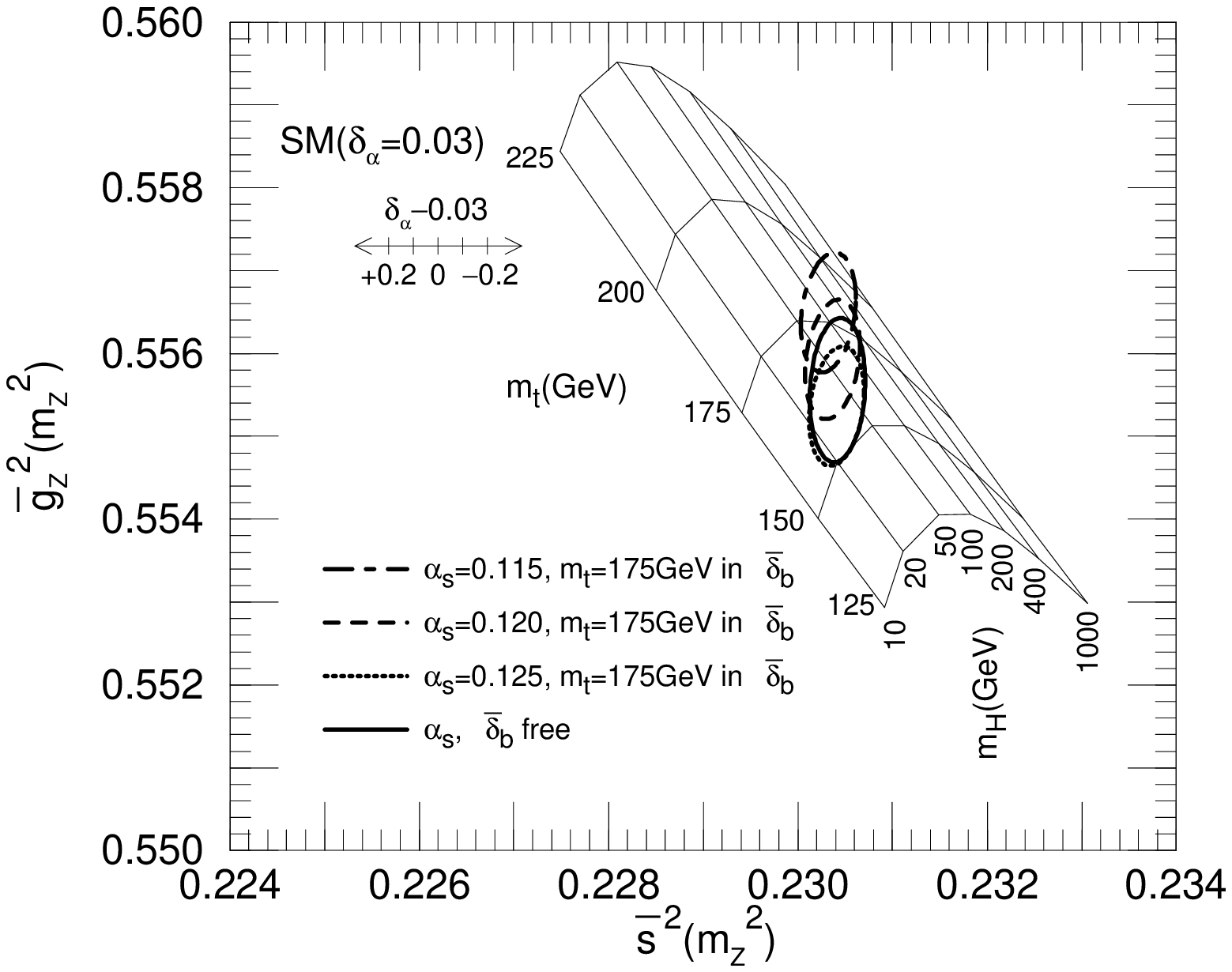,height=7cm,silent=0}
 \end{center}
\fcaption{%
A two-parameter fit to the $Z$ boson parameters in the
($\protect\sbar^2(\mmz), \protect\gzbar^2(\protect\mmz)$) plane,
where $\alpha_s(\protect\mz)$ is treated as an external
parameter and the $\protect\zbb$ vertex form-factor,
$\protect\delb(\protect\mmz)$, is evaluated in the SM
for $m_t=175$~GeV.
The 1-$\sigma$ (39\%CL) contours are shown for
$\alpha_s=0.115$, 0.120 and 0.125.
The solid contour is obtained by a four-parameter fit where
both $\alpha_s$ and $\delb(\mmz )$ are allowed to vary.
Also shown are the SM predictions in the range
$125\protect\gev\!<\!m_t\!<\!225\protect\gev$ and
$10\protect\gev\!<\!\protect\protect\mh\!<\!1000\protect\gev$ at
$\delta_\alpha \!\equiv\! 1/\bar{\alpha}(\protect\mmz)\!-\!128.72 \!=\!0.03$
and their dependences on $\delta_\alpha\!-\!0.03$.
}
\label{fig:gzb2sb2}
\end{figure}

The fit from the low-energy neutral-current data is
updated\cite{hhm96} by including the new CCFR data\cite{ccfr95_kevin}:
%
\fitoflenc
%
More discussions on the role of the low-energy neutral-current
experiments are given in the following subsection.

The $W$ mass data in Table~1,
$\mw = 80.26 \pm 0.16 \gev$,
gives
%
\fitofmw
for $\delg=0.0055$.

\begin{figure}[b]
 \begin{center}
 \leavevmode\psfig{file=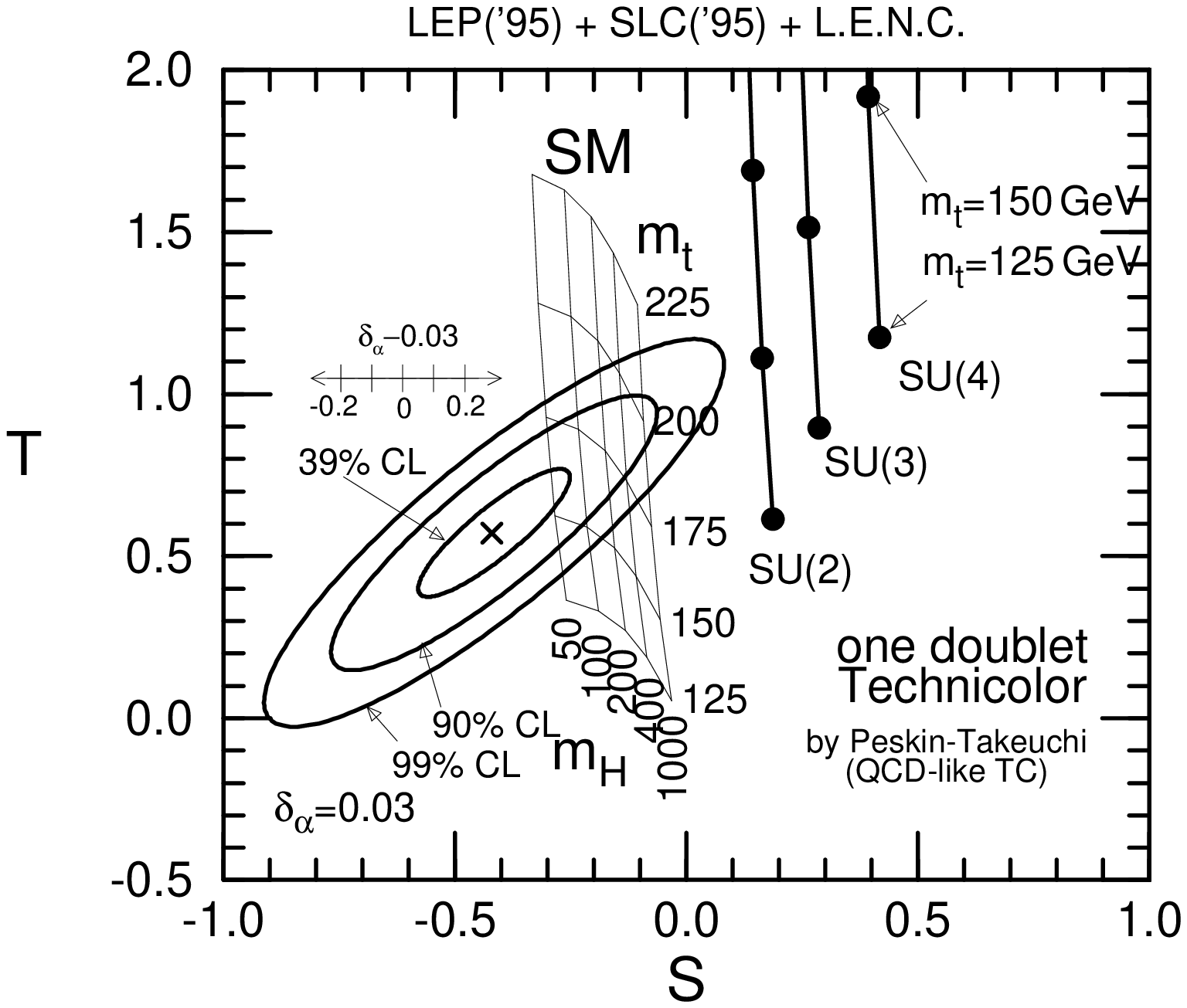,height=7cm,silent=0}
 \end{center}
\fcaption{%
Constraints on ($S$, $T$) from the five-parameter fit to
all the electroweak data for $\delta_\alpha=0.03$ and
$\protect\delg=0.0055$.
Together with $S$ and $T$, the $U$ parameter,
the $\protect\zbb$ vertex form-factor,
$\protect\delb(\protect\mmz)$, and the QCD coupling,
$\alpha_s(\mz)$, are allowed to vary in the fit.
Also shown are the SM predictions in the range
$125\protect\gev\!<\!m_t\!<\!225\protect\gev$
and $50\protect\gev\!<\!\protect\mh\!<\!1000\protect\gev$.
The predictions\protect\cite{stu} of one-doublet
${\protect\rm SU}(N_c)$--TC models are shown for $N_c=2,3,4$.
}
\label{fig:st}
\end{figure}

We perform a five-parameter fit to all the electroweak data,
the $Z$ parameters, the $W$ mass and the low-energy
neutral-current data,
in terms of $S$, $T$, $U$, $\delb$ and $\alpha_s$,
where we set $m_t=175$~GeV and $\mh=100$~GeV in the mild
running of the charge form-factors, e.g. in
Eq.(\ref{gzbarrunning}).
We find
%
\fitofstu
%
The dependence of the $S$ and $U$ parameters upon
$\delta_\alpha$ may be understood from Eq.(\ref{gbar_approx}).
For an arbitrary value of $\delg$ the parameter $T$ should be
replaced by $T'\!\equiv\! T\!+\!(0.0055\!-\!\delg)/\alpha$.
It should be noted that the uncertainty in $S$
coming from $\delta_\alpha=0.03\pm 0.09$ is
of the same order as that from the uncertainty in $\alpha_s$;
they are not negligible when compared to the overall error.
The $T$ parameter has little $\delta_\alpha$ dependence,
but it is sensitive to $\alpha_s$.

The above results, together with the SM predictions,
are shown in Fig.~8 as the projection onto the ($S,\,T$) plane.
Accurate parametrizations of the SM contributions to the
$S$, $T$, $U$ parameters are found in Ref.\cite{hhkm}.
Also shown are the predictions\cite{stu} of
the minimal (one-doublet)
SU($N_c$) Technicolor (TC) models with $N_c\!=\!2,3,4$.
It is clearly seen that the current experiments provide a
fairly stringent constraint on the simple TC models
if a QCD-like spectrum and the large $N_c$ scaling
are assumed\cite{stu}.
It is necessary for a realistic TC model to provide an
additional negative contribution to $S$\cite{appelquist93}
and a negligibly small contribution to $T$ at the same time.

Finally, if we regard the point $(S,T,U)=(0,0,0)$ as the
point with no-Electroweak corrections (a more precise
treatment will be given in section 5.2), then we find
$\chi^2_{\rm min}/({\rm d.o.f.})=141/(22)$ whose
probability less than $10^{-18}$.
On the other hand, if we also switch-off the remaining
electroweak corrections to $G_F$ by setting $\delg=0$,
then we find $T'=0.0055/\alpha=0.75$, and the point
$(S,T',U)=(0, 0.75, 0)$ gives
$\chi^2_{\rm min}/({\rm d.o.f.})=34.2/(22)$
which is consistent with the data at 5\%CL.
As emphasized in Ref.\cite{novikov93},
the genuine electroweak correction is not trivial to
establish in this analysis because of the cancellation
between the large $T$ parameter from $m_t \sim 175$~GeV and
the non-universal correction $\delg$ to the muon decay
constant in the observable combination\cite{hhkm}
$T'=T+(0.0055-\delg)/\alpha$.

\subsection{ Impact of the Low-Energy Neutral-Current Data }

In this subsection, we show individual contributions from the
four sectors of the low-energy neutral-current data\cite{hhkm},
$\nu_\mu$-$q$ and $\nu_\mu$-$e$ processes, atomic parity
violation (APV), and the classic $e$-$D$ polarization asymmetry
data.

The only new additional data this year is from the CCFR
collaboration\cite{ccfr95_kevin} which measured the ratio of
the neutral-current and charged-current cross-sections
in the $\nu_\mu$ scattering off nuclei.
By using the model-independent parameters of Ref.\cite{fh88},
they constrain the following linear combination,
\bea \label{k_ccfr}
K = 1.732 g_L^2
  + 1.119 g_R^2
  - 0.100 \delta_L^2
  - 0.086 \delta_R^2 \,,
\eea
and find
%
\dataofnqccfrorig
%
Because of the significantly high
$\langle Q^2 \rangle_{\rm CCFR} = 36{\gev}^2$
of the CCFR experiments as compared to the average of
the old data\cite{fh88}
($\langle Q^2 \rangle_{\rm HF} = 20{\gev}^2$),
the $q^2$-dependent electroweak corrections are different,
and we cannot combine the two data sets within the
model-independent framework.

\begin{figure}[t]
 \begin{center}
 \leavevmode\psfig{file=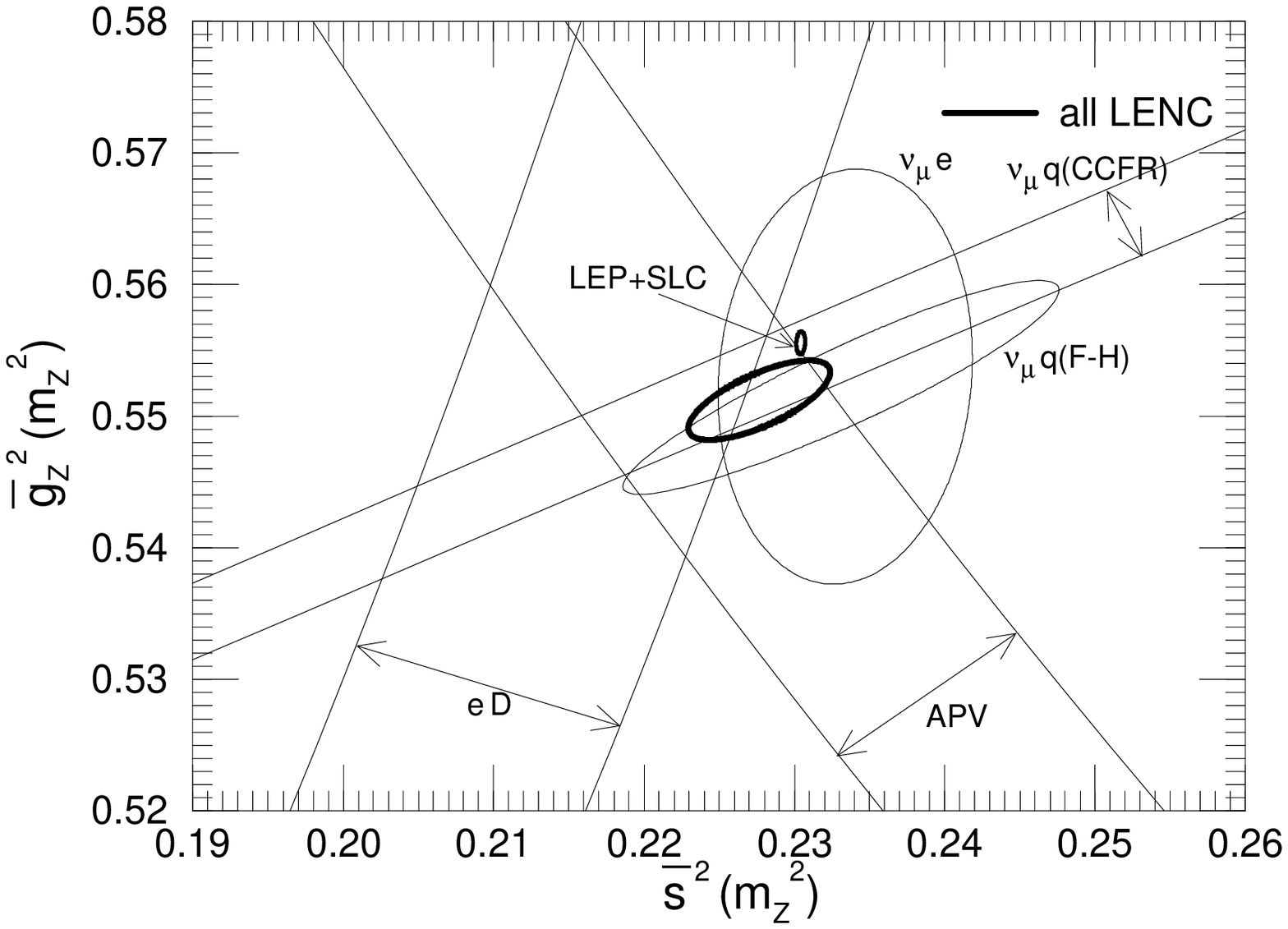,height=7cm,silent=0}
 \end{center}
\fcaption{%
Fit to the low-energy neutral-current data in terms of the
two universal charge form-factors $\protect\sbar^2(\mmz )$ and
$\protect\gzbar^2(\mmz )$.
1-$\sigma$ (39\%CL) contours are shown separately for
the old\cite{fh88} and the new\cite{ccfr95_kevin} $\nu_\mu$--$q$ data,
the $\nu_\mu$--$e$ data,
the atomic parity violation (APV) data,
and the SLAC $e$--${\rm D}$
polarization asymmetry data.
The 1-$\sigma$ contour of the combined fit,
Eq.(\protect\ref{fitoflencatmz}),
is shown by the thick contour.
Also shown is the constraint from the LEP/SLC data,
which is the solid contour in Fig.~7.
}
\label{fig:lenc}
\end{figure}

By noting that the data were obtained after correcting
for the external photonic corrections we find\cite{hhm96}
from the CCFR data (\ref{dataofnqccfrorig})
\fitofnqccfr
The corresponding fit to the old data\cite{fh88} gives\footnote{%
The data in Ref.\cite{fh88} were also corrected for the
external photonic corrections.
The $\delta_{c.c.}$ correction in Ref.\cite{hhkm} was
hence double counting.
The fit Eq.(4.17) of Ref.\cite{hhkm} has been revised here.}
\fitofnqfh

By combining with all the other neutral-current data of Ref.\cite{hhkm}
we find the fit Eq.(\ref{fitoflenc}).
In order to compare these constraints with those from the
LEP/SLC experiments it is useful to re-express the fit
in the $(\sbar^2(\mmz ),\gzbar^2(\mmz ))$ plane by
assuming the SM running of the charge form-factors.
The combined fit of Eq.(\ref{fitoflenc}) then becomes
%
\fitoflencatmz
%
In Fig.~9 we show individual contributions to the fit,
together with the combined LEP/SLC fit (the solid contour
of Fig.~7).
It is clear that the low-energy data have little impact
on constraining the effective charges, or equivalently
the $S$ and $T$ parameters.
They constrain, however, possible new interactions
beyond the ${\rm SU(2)_L\times U(1)_Y}$ gauge interactions,
such as those from an additional $Z$ boson\cite{langacker94}.
The model-independent parametrization of the low-energy
data is hence highly desirable.

\section{ The Minimal Standard Model Confronts the Electroweak Data }
In this section we assume that all the radiative corrections
are dominated by the SM contributions and obtain constraints
on $m_t$ and $\mh$ from the electroweak data.
%
\subsection{ Constraints on $m_t$ and $\mh$ as Functions of
		$\alpha_s$ and $\bar{\alpha}(\mmz)$ }
%
\begin{figure}[b]
 \begin{center}
  \leavevmode\psfig{file=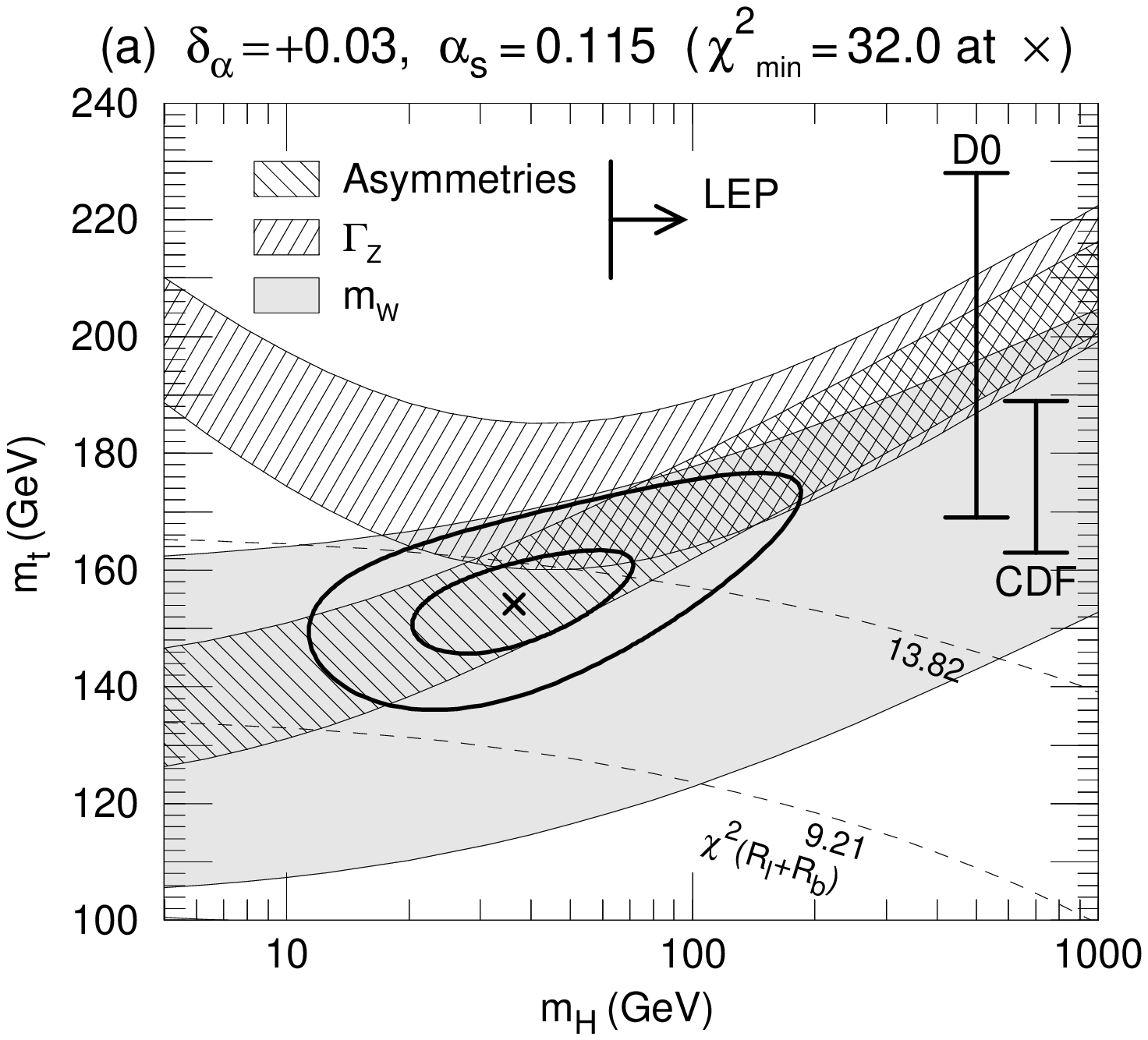,width=7.5cm,silent=0}
  \leavevmode\psfig{file=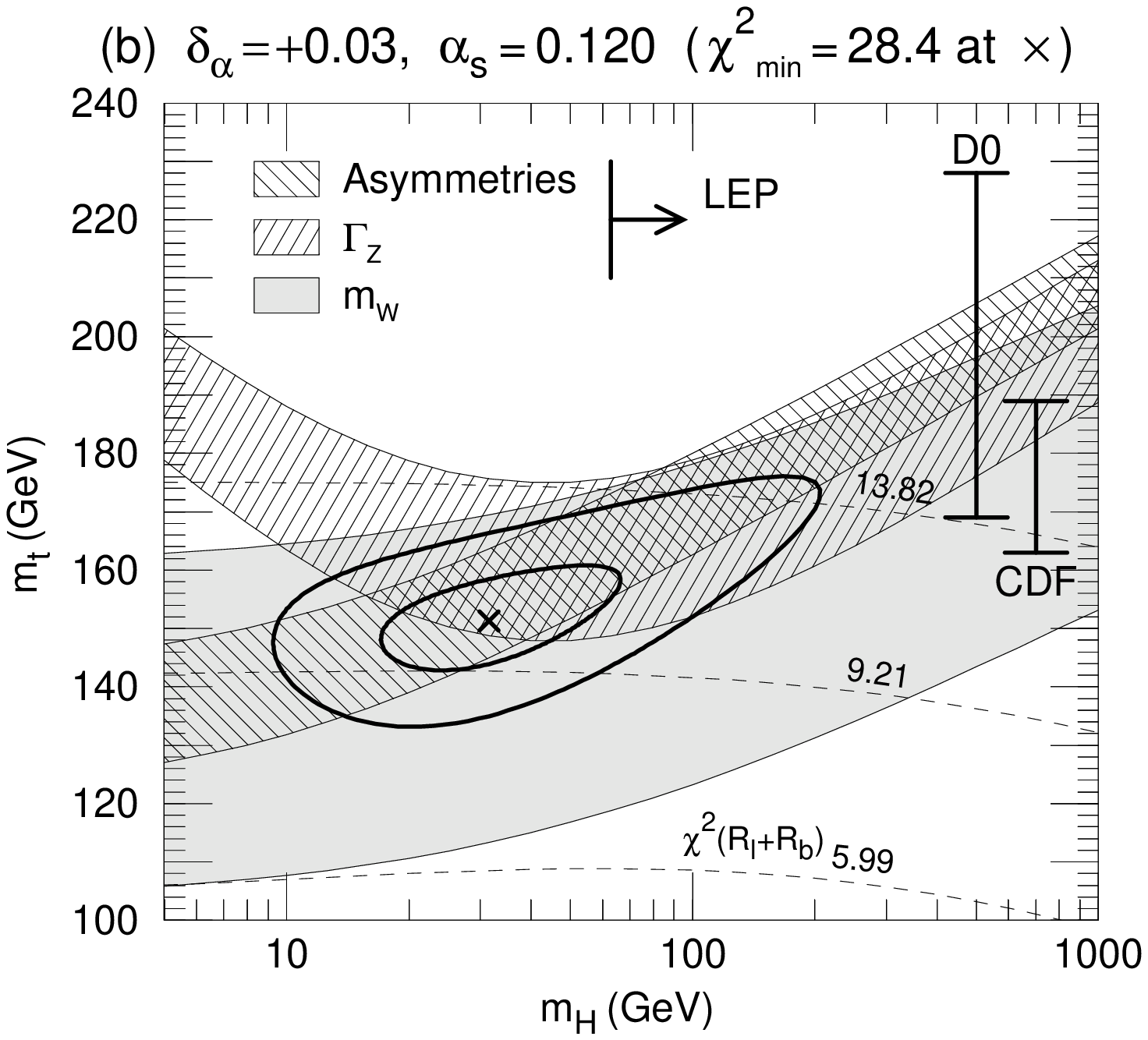,width=7.5cm,silent=0}\\[2mm]
  \leavevmode\psfig{file=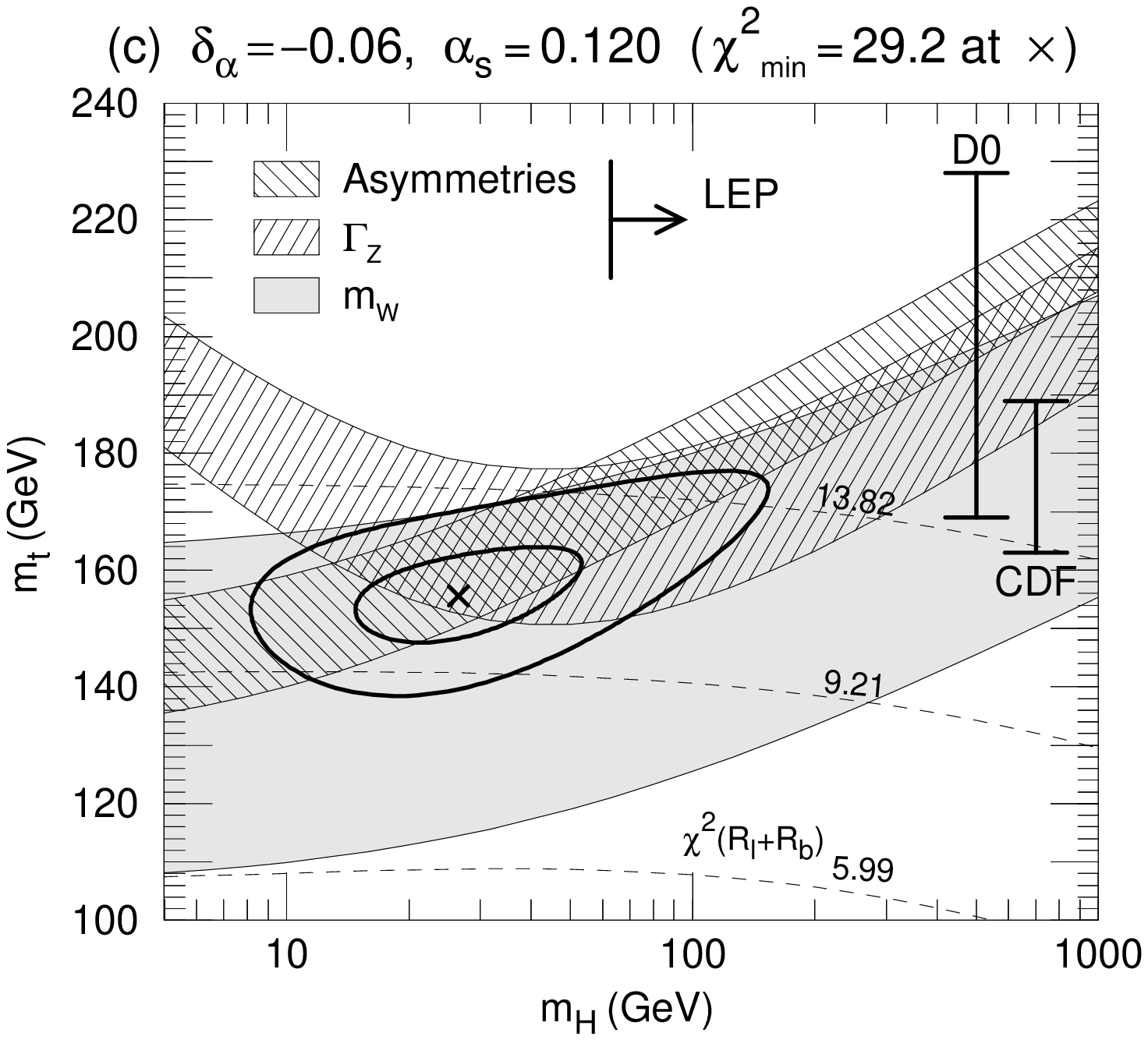,width=7.5cm,silent=0}
  \leavevmode\psfig{file=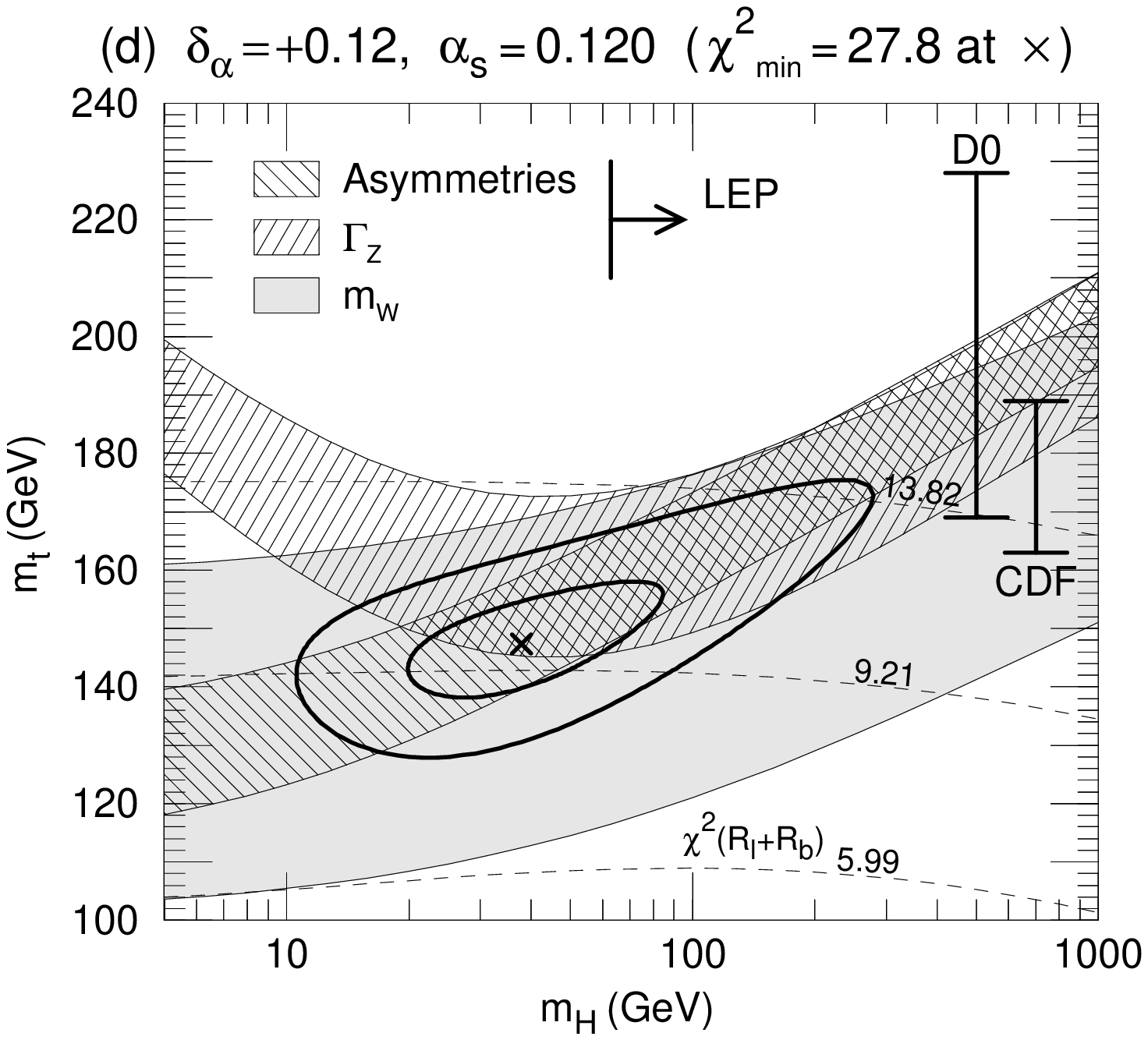,width=7.5cm,silent=0}
 \end{center}
\fcaption{%
The SM fit to all electroweak data in the ($\protect\mh,\,m_t$)
plane for
$(\delta_\alpha, \alpha_s)=(+0.03,0.115)$ (a),
$(+0.03,0.120)$ (b),
$(-0.06,0.120)$ (c) and
$(+0.12,0.120)$ (d),
where $\delta_\alpha=1/\bar{\alpha}(\mmz )-128.72$\cite{hhkm}.
The thick inner and outer contours correspond to
$\Delta\chi^2=1$ ($\sim$ 39\% CL),
and $\Delta\chi^2=4.61$ ($\sim$ 90\,\%~CL), respectively.
The minimum of $\chi^2$ is marked by an ``$\times$''.
Also shown are the 1-$\sigma$ constraints from the
$Z$-pole asymmetries, $\Gamma_Z$ and $\mw$.
The dashed lines show the constraint only from
$R_\ell$ and $R_b$.
The contours for $\chi^2\!=5.99,\,9.21,\,13.82$ correspond
to 95\%, 99\% and 99.9\%CL boundaries, respectively.
}
\label{fig:mtmh}
\end{figure}
%
In the minimal SM all the form-factors,
$\gzbar^2(\mmz)$, $\sbar^2(\mmz)$, $\gzbar^2(0)$, $\sbar^2(0)$,
$\gwbar^2(0)$ and $\delb(\mmz)$, depend uniquely
on the two mass parameters $m_t$ and $\mh$.
Fig.~10 shows the result of the global fit to all electroweak
data in the ($\mh,\,m_t$) plane for
(a) $\alpha_s=0.115$ and (b) 0.120 with $\delta_\alpha=0.03$,
and with (c) $\delta_\alpha\!=\!-0.06$
and (d) $+0.12$ for $\alpha_s=0.120$.
The thick inner and outer contours correspond to
$\Delta\chi^2\equiv\chi^2-\chi^2_{\rm min}=1$
(39\%CL),
and $\Delta\chi^2=4.61$ (90\%CL), respectively.
The minimum of $\chi^2$ is
indicated by an ``$\times$'' and the corresponding values
of $\chi^2_{\rm min}$ are given.
We also give the separate 1-$\sigma$ constraints arising from
the $Z$-pole asymmetries, $\Gamma_Z$, and $\mw$.
The asymmetries constrain $m_t$ and $\mh$ through
$\sbar^2(\mmz)$, while $\Gamma_Z$ constrains them through
the three form-factors $\gzbar^2(\mmz)$, $\sbar^2(\mmz)$ and
$\delb(\mmz)$.
In other words, the asymmetries measure the combination of
$S$ and $T$ as in Eq.(\ref{sbar_approx});
both $S$ and $T$ are functions of $m_t$ and $\mh$\cite{hhkm}.
On the other hand, $\Gamma_Z$ measures a different combination
of $S$ and $T$ with an additional constraint from $\delb$.
A remarkable point apparent from Fig.~10
is that, in the SM, when $m_t$ and $\mh$ are much larger than
$\mz$, $\Gamma_Z$ depends upon almost the same combination of
$m_t$ and $\mh$ as the one measured through $\sbar^2(\mmz)$.
This is because the quadratic $m_t$-dependence of
$\gzbar^2(\mmz)$ and that of $\delb$ largely cancel in
the SM prediction for $\Gamma_Z$.
Because of this only a band of $m_t$ and $\mh$ can be
strongly constrained from the asymmetries and $\Gamma_Z$ alone
despite their very small experimental errors.
The constraint from the $\mw$ data overlaps this allowed region.

\begin{figure}[t]
\begin{center}
 \leavevmode\psfig{file=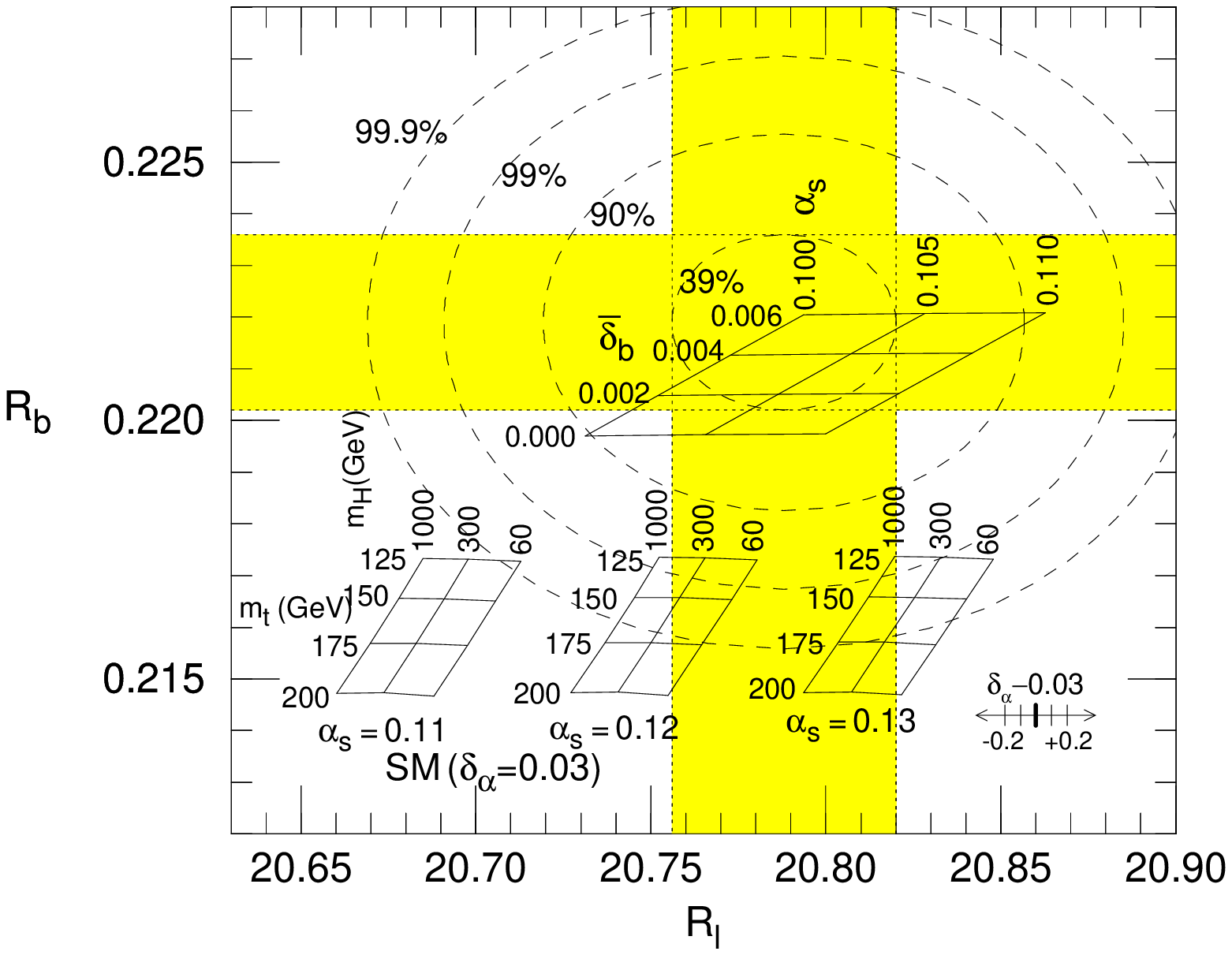,height=7cm,silent=0}
\end{center}
\fcaption{%
The $R_b$ vs $R_\ell$ plane.
The SM predictions are shown in the range
$120\protect\gev\!<\!m_t\!<\!240\gev$,
and $60\protect\gev\!<\!\protect\mh\!<\!1\tev$,
for three cases of $\alpha_s$
($\alpha_s$=$0.11$, $0.12$ and $0.13$).
These predictions are for $\delta_\alpha=0.03$, and
their dependences on $\delta_\alpha$ are also indicated.
Also shown are the 39\%,90\%,99\% and 99.9\%CL
contours obtained by combining only
the $R_\ell$ and $R_b$ data.
The ($\alpha_s,\delb$) lattice is obtained
by allowing $\delb$ to vary at
$\sbar^2(\mmz)=0.23039$ from Eq.(\ref{sb2_fit95}).
}
\label{fig:rlrb}
\end{figure}

Quantities which help to disentangle
the above $m_t$-$\mh$ correlation are $R_\ell$ and $R_b$.
The constraints from these data are shown in
Fig.~10 by dashed lines corresponding to
$\chi^2=5.99$ (95\%CL),
$\chi^2=9.21$ (99\%CL) and
$\chi^2=13.82$ (99.9\%CL) contours.
These constraints can be clearly seen in Fig.~11 where we
show the data and the SM predictions for $R_\ell$ and $R_b$.
$R_\ell$ is sensitive to the assumed value of $\alpha_s$, and,
for $\alpha_s=0.120$, the data favors smaller $\mh$.
$R_b$ is, on the other hand, sensitive to neither $\alpha_s$
nor $\mh$, and the data strongly disfavors large $m_t$.
It is thus the $R_\ell$ and $R_b$ data that constrain
the values of $m_t$ and $\mh$ from above.
If it were not for the data on $R_\ell$ and $R_b$
the common shaded region in Fig.~10 with very large $\mh$
$(\mh\sim 1\tev)$ could not be excluded by the
electroweak data alone.

It is clearly seen from Fig.~10 that the narrow ``asymmetry''
band is sensitive to $\delta_\alpha$, whereas the ``$\Gamma_Z$''
constraint is sensitive to $\alpha_s$.
The fit improves at larger $\delta_\alpha$ (larger
$1/\bar{\alpha}(\mmz )$) because the ``asymmetry'' constraint
then favors lower $m_t$ that is favored by the $R_b$ data.

The $\chi^2$ function of the global fit to all electroweak
data can be parametrized in terms of the four parameters
$m_t$, $\mh$, $\alpha_s$ and $\delta_\alpha$ :
%
\chisqsm
%
with
%
\fitofmt
%
and
%
\chisqhsm
%
Here $m_t$ and $\mh$ are measured in GeV.
This parametrization reproduces the exact $\chi^2$ function
within a few percent accuracy in the range
$100\gev<m_t<250\gev$,
$60\gev<\mh<1000\gev$ and
$0.10<\alpha_s(\mz)<0.13$.
The best-fit value of $m_t$ for a given set of $\mh$,
$\alpha_s$ and $\delta_\alpha$ is readily obtained from
Eq.(\ref{fitofmtbest}) with its approximate error of
(\ref{fitofmterror}).

For $\mh=60,300,1000\gev$, $\alpha_s=0.120\pm 0.07$ and
$\delta_\alpha=0.03\pm 0.09$, one obtains
\bea \label{mtfit_standard}
     m_t = 179 \pm 7 {}^{+19(\mh=1000)}_{-22(\mh=60)}
		\mp2 (\alpha_s)
		\mp5 (\delta_\alpha) \,,
\eea
where the mean value is for $\mh=300$~GeV.
The fit (\ref{mtfit_standard}) agrees excellently with
the estimate\cite{kleinknecht95}
\bea \label{mt_tevatron}
	m_t = 180 \pm 13~\gev
\eea
from the direct production data at Tevatron\cite{mt_cdf,mt_d0}.
Despite the claim\cite{novikov93} that there is no strong
evidence for the genuine electroweak correction, which we
re-confirmed in the previous section with the new data,
I believe that this is a strong evidence that the standard
electroweak gauge theory is valid at the quantum level.
The accidental cancellation of the two large radiative effects
in the observable combination $T'=T+(0.0055-\delg)/\alpha$
should give us, in the face of the Tevatron results
(\ref{mt_tevatron}), a strong evidence for the presence
of the electroweak correction to the muon decay, $\delg$,
which is finite and calculable only in the gauge
theory\cite{sirlin94}.

Due to the quadratic form of Eq.(\ref{total_chisqsm})
it is easy to obtain results which are independent of
$\alpha_s$ and/or $\delta_\alpha$.
Also, additional constraints on the external parameters
$\alpha_s$ and $\delta_\alpha$, such as those from their
improved measurements or the constraint from
the grand unification of these couplings
may be added without difficulty.
Here we give a parametrization of the constraint on
$\alpha_s\equiv \alpha_s(\mz )_{\msbar}$ from the electroweak data
within the minimal SM:
\bea \label{alps_ew}
\alpha_s = 0.1282 \pm 0.0035
	- 0.0105(\frac{m_t}{175})^2
	+ 0.00045\ln^2\frac{\mh}{7.8}
	- 0.0008\frac{\delta_\alpha-0.03}{0.09}\,,
\eea
which reproduces the results (see Fig.~4) well in the range
$150<m_t(\gev )<200$, $60<\mh(\gev )<1000$ and $|\delta_\alpha|<0.2$.

\begin{table}[t]
\begin{center}
\tcaption{ 95\%CL upper and lower bounds of $\mh$(GeV)
for a given $\alpha_s$ and
$\delta_\alpha=0.03\pm0.09$\cite{eidjeg95} }
\vspace*{1mm}

{\small
 \begin{tabular}{|c|clc|clc|clc|}
\hline
$\alpha_s$ &
\multicolumn{3}{|c|}{all EW data} &
\multicolumn{3}{|c|}{$-(R_b,R_c)$ data} &
\multicolumn{3}{|c|}{$+m_t$ (Tevatron)} \\
\hline
0.115 && $16<\mh<150$ &&& $18<\mh<290 $ &&& $22<\mh<360$ &\\ \hline
0.120 && $13<\mh<180$ &&& $15<\mh<500 $ &&& $20<\mh<550$ &\\ \hline
0.125 && $11<\mh<220$ &&& $12<\mh<1800$ &&& $18<\mh<980$ &\\ \hline
\end{tabular}
}

\end{center}
\end{table}

\begin{figure}[b]
\begin{center}
  \leavevmode\psfig{file=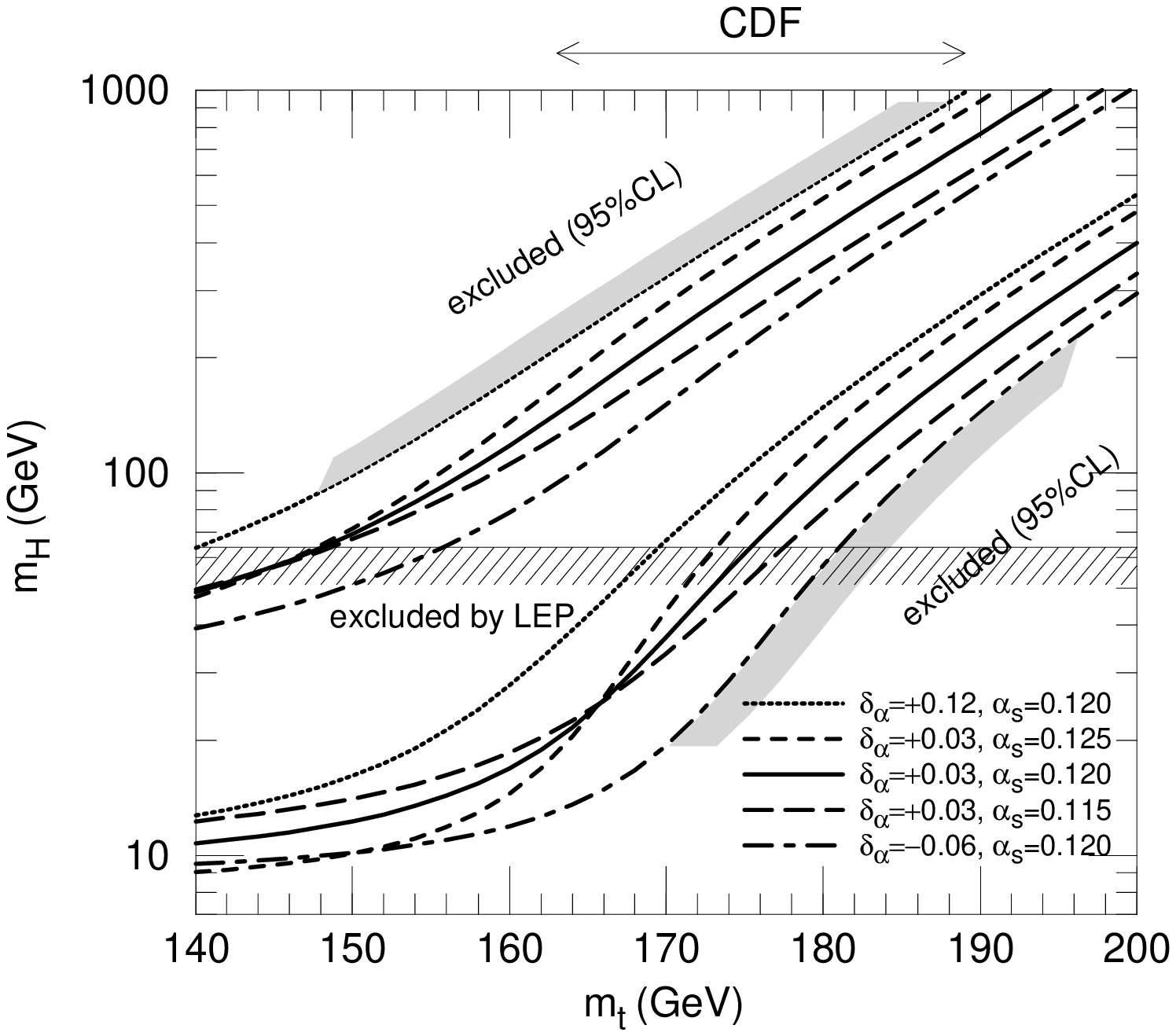,height=7cm,silent=0}
\end{center}
\fcaption{%
Constraints on the Higgs mass in the SM from all the
electroweak data.
Upper and lower bounds of the Higgs mass at 95\%~CL are
shown as functions of the top mass $m_t$, where
$m_t$ is treated as an external parameter
with negligible uncertainty.
The results are shown for $\alpha_s=0.120\pm 0.005$ and
$\delta_\alpha\equiv 1/\bar{\alpha}(\mmz )-128.72=0.03\pm 0.09$.
}
\label{fig:mhlimit}
\end{figure}

As discussed above, the constraint on $\mh$ from the electroweak
data is sensitive to the $R_b$, and hence on $\alpha_s$.
Shown in Table 5 are the 95\%CL upper and lower bounds on
$\mh$(GeV) from the electroweak data.
Low mass Higgs boson is clearly favored.
However, this trend disappears for $\alpha_s>0.12$
once we remove the $R_b$ and $R_c$ data.
The present $m_t$ estimate (\ref{mt_tevatron}) from
Tevatron does not significantly improve the situation.

\begin{figure}[b]
\begin{center}
\unitlength 0.9mm
\newcommand{\region}[1]{$\left.\vphantom{\line(0,1){#1}}\right\}$}
\newcommand{\mb}{\makebox}
\begin{picture}(100,115)(0,-5)
{\footnotesize
\put(0,  0.0){\line(0,1){100}}
\put(40, 0.0){\line(0,1){100}}
\put(0,  0.0){\line(1,0){40}}
\put(0, 14.1){\line(1,0){40}}
\put(0, 20.5){\line(1,0){40}}
\put(0, 27.9){\line(1,0){40}}
\put(0, 58.8){\line(1,0){40}}
\put(0, 63.3){\line(1,0){40}}
\put(0, 88.0){\line(1,0){40}}
\put(0,100.0){\line(1,0){40}}
\put( 20,106   ){\mb(0,0){range ($\sqrt{s}$)}}
\put( 20, 94   ){\mb(0,0){$\rho$}}
\put( 20, 75.65){\mb(0,0){1.05--2.5 GeV}}
\put( 20, 61.05){\mb(0,0){narrow resonances}}
\put( 20, 43.35){\mb(0,0){2.5--5 GeV}}
\put( 20, 24.2 ){\mb(0,0){5--7 GeV}}
\put( 20, 17.3 ){\mb(0,0){7--12 GeV}}
\put( 20,  7.05){\mb(0,0){$>$12 GeV}}
\put( 50,106   ){\mb(0,0)[l]{possible improvements in the future}}
\put( 67, 91   ){\mb(0,0)[l]{\region{11} $\tau$-charm \& B factories}}
\put( 67, 87   ){\mb(0,0)[l]{$\quad$(hadronic $\tau$-decays $+$ CVC)}}
\put( 43, 87   ){\mb(0,0)[l]{\region{10} DA$\Phi$NE}}
\put( 43, 83   ){\mb(0,0)[l]{$\quad$ VEPP-2M}}
\put( 50, 43.5 ){\mb(0,0)[l]{\region{20} $\tau$-charm factory}}
\put( 50, 17.4 ){\mb(0,0)[l]{\region{6} B factory}}
}
\end{picture}
\fcaption{%
Relative contributions to the uncertainty in
$\Delta\alpha_{\rm had}(\mmz)$ by
Burkhardt and Pietrzyk\cite{bp95}.
Possible improvements from the future generation
experiments are indicated.
}
\label{fig:delta_had_future}
\end{center}
\end{figure}

It is instructive to anticipate the impact a precise
measurement of the top-quark mass would have in the context of
the present electroweak data.
For instance, precision measurement of $m_t$ with an error of
1~GeV is envisaged at TeV33\cite{tev33}, a proposed luminosity
upgrade of Tevatron.
In the discussion below we treat $m_t$ as an external parameter,
and hence we discuss the sensitivity of the present electroweak
data to $\mh$ while assuming that $m_t$ is known precisely.

The 95\%CL upper/lower bounds on $\mh$ from the electroweak
data are shown in Fig.~12 as functions of $m_t$.
Dependences of the bounds on the two remaining parameters,
$\alpha_s=\alpha_s(\mz )_{\msbar}$ and
$\delta_\alpha=1/\bar{\alpha}(\mmz )-128.72$,
are shown clearly.
For a smaller value of $m_t$, $m_t<170$~GeV,
a rather stringent upper bound on $\mh$ is obtained,
whereas medium heavy Higgs boson is favored for
$m_t>180$~GeV.
It is tantalizing that the present data from Tevatron
(\ref{mt_tevatron}) lies just on the boundary.

Fig.~12 shows us that once the top-quark mass
is determined, either by direct measurements or by a
theoretical model, the major remaining uncertainty
is in $\delta_\alpha$, the magnitude of the QED
running coupling constant at the $\mz$ scale.
It is clear that we won't be
able to learn about $\mh$ in the SM, nor about
physics beyond the SM from its quantum effects,
without a significantly improved determination of
$\bar{\alpha}(\mmz )$.

Shown in Fig.~13 is the relative contributions to
the uncertainty in the present determination of
$\Delta\alpha(\mmz )$ based on the dispersion integral
over the $\sigma(e^+e^- \to {\rm hadrons})$ data,
taken from Ref.\cite{bp95}.
It is clearly seen that the majority of the uncertainty
comes from the low energy region, $\sqrt{s}<5$~GeV.
The $\phi$ factory DA$\Phi$NE and upgraded VEPP-2M
will be able to improve our knowledge at
$\sqrt{s} \simlt 1.5$~GeV.
However, by far the largest uncertainty comes from
the region $\sqrt{s}=2.5$--5.0~GeV which a future
$\tau$-charm factory can cover.
Precision measurements of the $\tau$ hadronic decay
rates will further give us normalization of the
$e^+e^-$ hadro-production cross section
upon use of the CVC (Conserved Vector Current)
rule of QCD\cite{dela_cvc}.
I believe that a $\tau$-charm factory, if realised in
the near future, will contribute most efficiently
toward sharpening of $\bar{\alpha}(\mmz )$.

\subsection{Is there already indirect evidence for the
standard $W$ self-coupling?}

The success of the SM predictions against precision electroweak
experiments at the quantum level suggests a question if there
is already evidence for the standard universal gauge-boson
self-couplings.
It is not trivial to answer this question definitely since
we should identify which finite portion of the quantum
corrections is sensitive to the weak-boson self-interactions.
Usually one splits the complete SM radiative corrections into
just two pieces which are separately gauge invariant,
the fermionic loop contributions to the gauge-boson
self-energies, and the rest.
It can then be stated clearly that neither of the corrections
alone is consistent with the data, and both contributions are
needed to explain the success of the SM radiative
corrections\cite{sirlin94}.
Since the bosonic part of the correction should necessarily
contain the weak boson self-interactions, we may already have
evidence for universal couplings.

\begin{table}[t]
\begin{center}
\tcaption{%
The electroweak data and the SM predictions.
The three predictions for $\Gamma_Z$, $\sigma_h^0$ and $R_\ell$
are for $\alpha_s=0.115$, 0.120 and 0.125.
}
 \def\equnit{$\times 10^{-5}$}
 \def\afb{A_{\rm FB}}
 \def\non{$\;\;$------}
{ \scriptsize
 \begin{tabular}{|l|r|l|l|l|l|l@{\,\,}l@{\,\,}l@{\,}|}
 \hline
  & data \hspace{7mm} & no-EW &
  +fermion  & +box& +vertex &
  \multicolumn{3}{|c|}{+propagator}\\
 \hline
         $m_t$ (GeV)   &
& \non  &    175 &    175 &    175 &    175 &    175 &    175\\
         $m_H$ (GeV)   &
& \non  &    100 &   \non &    100 &     60 &    300 &   1000\\
 \hline
$S$                    &
 &  \non  & -0.066 & -0.066 & -0.066 & -0.283 & -0.146 & -0.075\\
$T$                    &
 &  \non  &  1.134 &  1.134 &  1.134 &  0.908 &  0.759 &  0.578\\
$U$                    &
 &  \non  &  0.014 &  0.014 &  0.014 &  0.357 &  0.352 &  0.351\\
$\bar{\delta}_G$       &
 &  \non  & \non   &0.00429 &0.00549 &0.00549 &0.00549 &0.00549\\
$1/\bar{\alpha}(m_Z^2)$&
 & 128.89 & 128.90 & 128.90 & 128.90 & 128.75 & 128.75 & 128.75\\
$\bar{s}^2(m_Z^2)$     &
 &0.23112 &0.22814 &0.22954 &0.22994 &0.23008 &0.23093 &0.23162\\
$\bar{g}_Z^2(m_Z^2)$   &
 &0.54865 &0.55813 &0.55571 &0.55504 &0.55642 &0.55594 &0.55520\\
$\bar{\delta}_b(m_Z^2)$&
& \non & \non  & \non  &-0.00993 &-0.00995 &-0.00991 &-0.00997\\
$\bar{s}^2(0)$         &
 &0.23865 &0.23583 &0.23715 &0.23752 &0.23850 &0.23930 &0.23995\\
$\bar{g}_Z^2(0)$       &
 &0.54865 &0.55323 &0.55085 &0.55019 &0.54928 &0.54868 &0.54796\\
$\bar{g}_W^2(0)$       &
 &0.42185 &0.42714 &0.42453 &0.42380 &0.42448 &0.42338 &0.42236\\
 \hline
$\Gamma_Z$(GeV) &  2.4963 $\pm$ 0.0032
 & 2.4838 & 2.5348 & 2.5200 & 2.4907 & 2.4965 & 2.4922 & 2.4870\\
                &
 & 2.4866 & 2.5377 & 2.5229 & 2.4935 & 2.4994 & 2.4950 & 2.4898\\
                &
 & 2.4894 & 2.5405 & 2.5257 & 2.4963 & 2.5022 & 2.4978 & 2.4926\\
$\sigma_h^0$(nb)&  41.488 $\pm$  0.078
 & 41.505 & 41.498 & 41.500 & 41.487 & 41.487 & 41.490 & 41.494\\
                &
 & 41.478 & 41.471 & 41.474 & 41.460 & 41.461 & 41.464 & 41.468\\
                &
 & 41.452 & 41.445 & 41.447 & 41.434 & 41.434 & 41.437 & 41.441\\
$R_\ell$        &  20.788 $\pm$  0.032
 & 20.768 & 20.817 & 20.795 & 20.734 & 20.731 & 20.716 & 20.703\\
                &
 & 20.802 & 20.851 & 20.828 & 20.767 & 20.765 & 20.750 & 20.737\\
                &
 & 20.835 & 20.884 & 20.862 & 20.801 & 20.798 & 20.784 & 20.770\\
$\afb^{0,\ell}$ &  0.0172 $\pm$ 0.0012
 & 0.0169 & 0.0224 & 0.0198 & 0.0175 & 0.0172 & 0.0157 & 0.0145\\
$A_\tau$        &  0.1418 $\pm$ 0.0075
 & 0.1502 & 0.1733 & 0.1625 & 0.1517 & 0.1506 & 0.1439 & 0.1384\\
$A_e$           &  0.1390 $\pm$ 0.0089
 & 0.1502 & 0.1733 & 0.1625 & 0.1517 & 0.1506 & 0.1439 & 0.1384\\
$R_b$           &  0.2219 $\pm$ 0.0017
 & 0.2182 & 0.2181 & 0.2182 & 0.2157 & 0.2157 & 0.2157 & 0.2157\\
$R_c$           &  0.1540 $\pm$ 0.0074
 & 0.1717 & 0.1719 & 0.1718 & 0.1722 & 0.1722 & 0.1721 & 0.1721\\
$\afb^{0,b}$    &  0.0997 $\pm$ 0.0031
 & 0.1054 & 0.1219 & 0.1142 & 0.1063 & 0.1056 & 0.1008 & 0.0968\\
$\afb^{0,c}$    &  0.0729 $\pm$ 0.0058
 & 0.0753 & 0.0882 & 0.0822 & 0.0763 & 0.0757 & 0.0720 & 0.0690\\
  $\sin^2\theta^{lept}_{eff}(\langle Q_{\rm FB}\rangle)$
&  0.2325 $\pm$ 0.0013
 & 0.2311 & 0.2282 & 0.2296 & 0.2309 & 0.2311 & 0.2319 & 0.2326\\
$A_{\rm LR}$    &  0.1551 $\pm$ 0.0040
 & 0.1502 & 0.1733 & 0.1625 & 0.1517 & 0.1506 & 0.1439 & 0.1384\\
$A_b({\rm LR})$ &   0.841 $\pm$  0.053
 &  0.936 &  0.938 &  0.937 &  0.935 &  0.935 &  0.934 &  0.934\\
$A_c({\rm LR})$ &   0.606 $\pm$  0.090
 &  0.669 &  0.679 &  0.674 &  0.670 &  0.669 &  0.666 &  0.664\\
    $\chi^2$        &$(\alpha_s=0.115)\quad $
 &   35.0 &  296.4 &  117.2 &   34.5 &   30.2 &   34.4 &   57.1\\
    (d.o.f.=14)     &$(\alpha_s=0.120)\quad $
 &   28.7 &  320.3 &  131.8 &   29.6 &   28.2 &   29.3 &   48.4\\
                    &$(\alpha_s=0.125)\quad $
 &   26.6 &  348.3 &  150.5 &   28.8 &   30.4 &   28.3 &   43.9\\
 \hline
$g_L^2$         &  0.2980 $\pm$ 0.0044
 & 0.2955 & 0.3027 & 0.3049 & 0.3067 & 0.3049 & 0.3037 & 0.3024\\
$g_R^2$         &  0.0307 $\pm$ 0.0047
 & 0.0309 & 0.0307 & 0.0307 & 0.0298 & 0.0300 & 0.0301 & 0.0302\\
$\delta_L^2$    & -0.0589 $\pm$ 0.0237
 &-0.0601 &-0.0606 &-0.0652 &-0.0645 &-0.0645 &-0.0645 &-0.0645\\
$\delta_R^2$    &  0.0206 $\pm$ 0.0160
 & 0.0186 & 0.0184 & 0.0184 & 0.0179 & 0.0180 & 0.0180 & 0.0181\\
$\chi^2 $       &
 &    0.4 &    1.8 &    4.0 &    5.5 &    3.6 &    2.4 &    1.5\\
 \hline
$K$ (CCFR)      &  0.5626 $\pm$ 0.0060
 & 0.5519 & 0.5641 & 0.5685 & 0.5703 & 0.5674 & 0.5653 & 0.5632\\
$\chi^2 $       &
 &    3.2 &    0.1 &    1.0 &    1.6 &    0.6 &    0.2 &    0.0\\
 \hline
$s^2_{eff}$     &   0.233 $\pm$  0.008
 &  0.239 &  0.236 &  0.235 &  0.229 &  0.230 &  0.231 &  0.231\\
$\rho_{eff}$    &   1.007 $\pm$  0.028
 &  1.000 &  1.008 &  1.016 &  1.015 &  1.013 &  1.012 &  1.011\\
$\chi^2$        &
 &    0.6 &    0.1 &    0.1 &    0.4 &    0.2 &    0.1 &    0.1\\
 \hline
$Q_W$           &  -71.04 $\pm$   1.81
 & -74.73 & -74.73 & -72.96 & -72.91 & -73.00 & -73.10 & -73.14\\
$\chi^2$        &
 &    4.2 &    4.2 &    1.1 &    1.1 &    1.2 &    1.3 &    1.3\\
 \hline
$2C_{1u}-C_{1d}$&   0.938 $\pm$  0.264
 &  0.709 &  0.725 &  0.730 &  0.729 &  0.724 &  0.721 &  0.718\\
$2C_{2u}-C_{2d}$&  -0.659 $\pm$  1.228
 &  0.082 &  0.100 &  0.103 &  0.112 &  0.106 &  0.101 &  0.097\\
$\chi^2$        &
 &    1.9 &    1.2 &    1.1 &    1.1 &    1.2 &    1.4 &    1.5\\
 \hline
$m_W$           &   80.26 $\pm$   0.16
 &  79.96 &  80.46 &  80.38 &  80.36 &  80.43 &  80.32 &  80.23\\
$\chi^2$        &
 &    3.5 &    1.6 &    0.6 &    0.4 &    1.1 &    0.2 &    0.0\\
 \hline
  $\chi^2_{\rm tot}$&$(\alpha_s=0.115)\quad $
 &   48.8 &  305.3 &  125.0 &   44.6 &   38.1 &   40.0 &   61.6\\
  (d.o.f.=25)       &$(\alpha_s=0.120)\quad $
 &   42.5 &  329.2 &  139.6 &   39.7 &   36.1 &   34.9 &   52.9\\
                    &$(\alpha_s=0.125)\quad $
 &   40.4 &  357.2 &  158.4 &   38.9 &   38.3 &   33.9 &   48.3\\
 \hline
 \end{tabular}
}
\end{center}
\end{table}
%

It is not clear to me, however, how much of these finite
bosonic correction terms depend on the splitting of the
gauge bosons into themselves.
For instance, the box diagrams do not contain gauge-boson
self-couplings.
I therefore split the bosonic corrections into three separately
gauge-invariant pieces, `box-like', `vertex-like' and
`propagator-like' pieces by appealing to the S-matrix pinch
technique\cite{pinch}.
It is then only the `vertex-like' and `propagator-like' pieces
which contain the gauge boson self-couplings.
Schematically we separate the SM radiative corrections into
the following five pieces:
  \SetScale{0.2}
  \begin{eqnarray}
     \begin{array}{lll@{\quad\hspace{1mm}\quad}l}
      {\cal M}\quad=&
       \quad \mbox{QED/QCD} & & (A)
       \\[0.3mm]
      &\mbox{+ fermion-loop} &
       \begin{picture}(40,20)(2,8)
         \CArc(120,50)(40,0,180)
         \CArc(120,50)(40,180,360)
         \Photon(20,50)(80,50){7}{3}   
         \Photon(160,50)(220,50){7}{3} 
       \end{picture}
       & (B)
       \\[0.3mm]
      &\mbox{+ box} &
       \begin{picture}(40,20)(-12,8)
         \Line(0,0)(0,100)
         \Line(100,0)(100,100)
         \Photon(0,25)(100,25){7}{5}
         \Photon(0,75)(100,75){7}{5}
       \end{picture}
       & (C)
       \\[0.3mm]
       &\mbox{+ vertex} &
       \begin{picture}(40,20)(-12,8)
         \Line(30,0)(30,100)
         \PhotonArc(30,50)(30,90,270){7}{4}
         \Photon(30,50)(100,50){7}{4}
       \end{picture}
       +
       \begin{picture}(40,20)(-12,8)
         \Line(0,0)(0,100)
         \Photon(50,50)(100,50){7}{3}
         \Photon(0,20)(50,50){7}{3}
         \Photon(0,80)(50,50){7}{3}
       \end{picture}
       & (D)
       \\[0.3mm]
       &\mbox{+ bosonic-loop} &
       \begin{picture}(45,20)(2,8)
         \PhotonArc(120,50)(40,0,180){7}{6}
         \PhotonArc(120,50)(40,180,360){7}{6}
         \Photon(20,50)(80,50){7}{3}  \Vertex(80,50){1}
         \Photon(160,50)(220,50){7}{3} \Vertex(160,50){1}
       \end{picture}
       +
       \begin{picture}(40,20)(5,8)
         \PhotonArc(120,50)(40,0,360){7}{11}
         \Photon(36,0)(211,0){7}{9}
       \end{picture}
       +
       \begin{picture}(45,15)(2,8)
         \DashCArc(120,50)(40,0,180){10}
         \PhotonArc(120,50)(40,180,360){7}{6}
         \Photon(20,50)(80,50){7}{3}  \Vertex(80,50){1}
         \Photon(160,50)(220,50){7}{3} \Vertex(160,50){1}
       \end{picture}
     & (E)
     \end{array}\label{abcde}
  \end{eqnarray}
Details of this separation for each radiative correction term
may be obtained straightforwardly from the analytic expressions
presented in Ref.\cite{hhkm}.
We find by confronting these `predictions' with the latest
electroweak data the results of Table~6.

The `no-EW' entry confronts the tree-level predictions of the SM
where only QCD and external QED corrections are applied.
In this column $\bar{\alpha}(m_Z^2)$ is calculated by
including only contributions from light quarks and leptons
with $\delta_h=0.03$\cite{hhkm} for the hadronic uncertainty.
It is quite striking to re-confirm the
observation\cite{novikov93} that these `no-EW' predictions
agree with experiments at LEP/SLC very well.
In fact, it gives even better $\chi^2$ than the SM, partly
because of the $R_b$ data, which prefers no electroweak
corrections $\delb(\mmz )=0$ over the SM prediction
$\delb(\mmz )=-0.0099$ for $m_t=175$~GeV.
It is only the $m_W$ value\cite{novikov94} and the atomic
parity violation data which give significantly higher $\chi^2$
than the SM does.

The next `${\rm + fermion}$' column\footnote{%
$\mh=100$~GeV is chosen to fix the negligible two-loop
contributions in the `${\rm + fermion}$' and
`${\rm + vertex}$' columns.
}~gives the result of
$A+B$ in Eq.(\ref{abcde}).
That the LEP/SLC data can be fitted well by the `no-EW'
calculation is a consequence of an accidental cancellation
between the vertex/box correction to the $\mu$ decay matrix
elements (the factor $\delg$ in the Table) and
the $T$ parameter for $m_t \sim 175~\gev$
in the observable combination $T'\equiv T+(0.0055-\delg)/\alpha$.
If we include only the fermionic corrections
the $T$ parameter grows from zero to 1.144, while the factor
$\delg$ remains zero \cite{hhkm}.
The jump of $\chi^2$ from 30 to 300 in the LEP/SLC experiments
is a consequence of the absence of this cancellation in $T'$.

It turned out that the `box-like' corrections to the
$\mu$-decay matrix elements give almost 80\% of the total
$\delg$ value.
Hence by adding the `box-like' corrections,
the fit improves significantly.
This can be seen from the column of `${\rm + box}$',
where we give results of A+B+C corrections in Eq.(\ref{abcde}).

Up to this stage no contribution from quantum fluctuations
with the weak-boson self-couplings are counted.
It is in the next step, the `${\rm + vertex}$' column
where I list the results of A+B+C+D corrections,
we can start to see their effects.
It turns out that the effects of the remaining 20\% correction
to $\delg$ and the effects in part from the vertex
corrections in the $Z$-decay matrix elements significantly
reduce the $\chi^2$ in the LEP/SLC sector of the experiments
from about 130 down to 30.

I should therefore conclude that the effect of the
`vertex-like' corrections is significant for the success of
the SM at the quantum correction level.
Even setting aside the fundamental problem that we could not
control quantum fluctuations at short distances if it were not
for the universality of the weak-boson self-couplings,
it is reassuring to learn that, after cancellation of the
short-distance singularity, the remaining finite correction
makes the fit even better.
I note in passing that the significance of the
`propagator-like' correction term which contains the Higgs-mass
dependence of the SM prediction cannot be established at the
present level of accuracy.

\section{Conclusions}

(i) 	The precision electroweak experiments at LEP and SLC test
the SM predictions at a few times $10^{-3}$ level, which is sufficient
to resolve some of the radiative effects.

(ii)	All the data agree well with the predictions of the SM except for
$R_b$ and $R_c$ measured at LEP, which gives 3\% larger $R_b$ at
3.7-$\sigma$ and 11\% smaller $R_c$ at 2.5-$\sigma$.  When combined
the two data alone would rule out the SM at 99.99\%CL for
$m_t>170$~GeV.

(iii)	If we allow only $\Gamma_b$ and $\Gamma_c$ to deviate from the
SM predictions, then the data on $R_b$ and $R_c$ implies un-acceptably
large $\alpha_s$.

(iv)	If we assume the SM value for $\Gamma_c$, then the $R_b$ data
is still 2\% larger than the SM prediction at 3-$\sigma$.
Several theoretical models have been proposed to explain the discrepancy.
The common consequence of allowing only $\Gamma_b$ to deviate from
the SM is small $\alpha_s$, $\alpha_s=0.104\pm 0.008$.

(v)	If we allow the QCD coupling $\alpha_s$ to vary in the global fit
to the electroweak data, the $R_b$ problem does not affect the standard
$S$, $T$, $U$ analysis.  The ($S$, $T$) fit agrees excellently with
the SM, but disfavors the naive QCD-like technicolor models.

(vi)	The global fit in the minimal SM constrains ($m_t$, $\mh$),
where the prefered $m_t$ range agrees well with the top-quark data at
Tevatron.

(vii)	Once $m_t$ is known precisely, an improved constraint on $\mh$
from precision electroweak experiments will be achieved only with the
improved measurement on $\Delta\alpha_{\rm had}(\mmz )$.
The contribution of a future $\tau$-charm factory will be decisive.

(viii)	The agreement of the SM predictions with precision experiments
improves significantly when one includes radiative effects due to
`vertex-like' corrections which may be regarded as indirect evidence
for the universal weak-boson self-couplings.

\section{Acknowledgements}
I would like to thank
D.~Haidt, N.~Kitazawa, S.~Matsumoto and Y.~Yamada
for fruitful collaborations which made this presentation possible.
I would also like to thank
S.~Aoki, B.~Bullock, D.~Charlton, M.~Drees, S.~Erredi,
G.L.~Fogli, R.~Jones, J.~Kanzaki,
C.~Mariotti, A.D.~Martin, K.~McFarland,
T.~Mori, M.~Morii, D.R.O.~Morrison,
B.~Pietrzyk, P.B.~Renton,
M.H.~Shaevitz, D.~Schaile, M.~Swartz, R.~Szalapski,
T.~Takeuchi, P.~Vogel, P.~Wells and D.~Zeppenfeld
for discussions that helped me understand the experimental data
and their theoretical implications better.


\section{References}

\end{document}